\documentclass[useAMS,usegraphicx]{mn2e}
\usepackage{epsfig}
\newcommand{\mum}   {$\mu$m}
\newcommand{\kms}   {km~s$^{-1}$}

\newcommand{\cmt}   {cm$^{-3}$}
\newcommand{\jpb}   {$\rm Jy~beam^{-1}$}    
\newcommand{\lo}    {$L_{\sun}$}
\newcommand{\mo}    {$M_{\sun}$}
\newcommand{\mj}    {$M_\mathrm{Jup}$}
\newcommand{\co}    {$^{12}$CO}
\newcommand{\tco}    {$^{13}$CO}

\newcommand{\nth}   {N$_2$H$^+$}

\newcommand{\water}  {H$_2$O}

\newcommand{\et}    {et al.}
\newcommand{\eg}    {e.\,g.,}
\newcommand{\ie}    {i.\,e.,}

\newcommand{\supa}  {$^\mathrm{a}$}
\newcommand{\supb}  {$^\mathrm{b}$}
\newcommand{\supc}  {$^\mathrm{c}$}
\newcommand{\supd}  {$^\mathrm{d}$}
\newcommand{\supe}  {$^\mathrm{e}$}

\newcommand{\phe}   {\phantom{$^\mathrm{c}$}}
\newcommand{\phb}   {\phantom{$>$}}

\newcommand{\phda}  {\phantom{\dag}}
\title[A search for proto-BDs in Taurus]{A search for pre-substellar cores and proto-brown dwarf candidates in Taurus: multiwavelength analysis in the B213-L1495 clouds 
}
\author[Palau et al.]{Aina Palau$^{1}$\thanks{E-mail:palau@ieec.uab.es}, I. de Gregorio-Monsalvo$^{2,3}$, \`O. Morata$^{4}$, D. Stamatellos$^5$, N. Hu\'elamo$^6$
\newauthor
C. Eiroa$^{7}$, A. Bayo$^2$, M. Morales-Calder\'on$^{8}$, H. Bouy$^6$, \'A. Ribas$^{9}$, D. Asmus$^{10}$,
\newauthor
D. Barrado$^{6,11}$
\\
\\
$^{1}$ Institut de Ci\`encies de l'Espai (CSIC/IEEC), Campus UAB, Facultat de Ci\`encies, Torre C-5 parell 2, E-08193 Bellaterra, Spain\\
$^{2}$ European Southern Observatory, Alonso de C\'ordova 3107, Vitacura, Santiago, Chile\\
$^{3}$ Joint ALMA Office, Alonso de C\'ordova 3107, Vitacura, Santiago, Chile\\
$^{4}$ Academia Sinica, Institute of Astronomy and Astrophysics, P.O. Box 23-141, Taipei  106, Taiwan\\
$^{5}$ School of Physics \& Astronomy, Cardiff University,  5 The Parade, Cardiff CF24 3AA, UK\\
$^{6}$ Dpto. Astrof\'{\i}sica, Centro de Astrobiolog\'{\i}a (INTA-CSIC), ESAC Campus, PO Box 78, E-28691 Villanueva de la Ca\~nada, Spain\\
$^{7}$ Dpto. F\'{\i}sica Te\'orica,  Facultad de Ciencias, Universidad Aut\'onoma de Madrid,  E-28049 Madrid, Spain\\
$^{8}$ Spitzer Science Center, California Institute of Technology, 1200 E California Blvd., Pasadena, CA 91125, USA\\
$^{9}$ Herschel Science Centre, European Space Astronomy Centre (ESA), P.O. Box 78, E-28691 Villanueva de la Ca\~nada, Madrid, Spain\\
$^{10}$ Institut f\"ur Theoretische Physik und Astrophysik, Christian-Albrechts-Universit\"at zu Kiel, Leibnizstr. 15, 24098, Kiel, Germany\\
$^{11}$ Centro Astron\'omico Hispano Alem\'an de Calar Alto (CAHA), C/ Jes\'us Durb\'an Rem\'on 2-2, E-04004 Almer\'{\i}a, Spain\\
}
\begin{document}

\date{Accepted date. Received date; in original form date}

\pagerange{\pageref{firstpage}--\pageref{lastpage}} \pubyear{2012}

\maketitle

\label{firstpage}

\begin{abstract}
In an attempt to study whether the formation of brown dwarfs (BDs) takes place as a scaled-down version of low-mass stars,
we conducted IRAM\,30\,m/MAMBO-II observations at 1.2~mm in a sample of 12 proto-BD candidates selected from Spitzer/IRAC data in the B213-L1495 clouds in Taurus. Subsequent observations with the CSO at 350~$\mu$m, VLA at 3.6 and 6 cm, and IRAM\,30\,m/EMIR in the $^{12}$CO\,(1--0), $^{13}$CO\,(1--0), and N$_2$H$^+$\,(1--0) transitions were carried out toward the two most promising Spitzer/IRAC source(s), J042118 and J041757.
J042118 is associated with a compact ($<10$~arcsec or $<1400$~AU) and faint source at 350~$\mu$m, while J041757 is associated with a partially resolved ($\sim16$~arcsec or $\sim2000$~AU) and stronger source emitting at centimetre wavelengths with a flat spectral index. The corresponding masses of the dust condensations are $\sim1$ and $\sim5$~$M_\mathrm{Jup}$ for J042118 and J041757, respectively. In addition, about 40~arcsec to the northeast of J041757 we detect a strong and extended submillimetre source, J041757-NE,  which is not associated with NIR/FIR emission down to our detection limits, but is clearly detected in $^{13}$CO and N$_2$H$^+$ at $\sim7$~km~s$^{-1}$, and for which we estimated a total mass of $\sim100$~$M_\mathrm{Jup}$, close to the mass required to be gravitationally bound. In summary, our observational strategy has allowed us to find in B213-L1495 two proto-BD candidates and one pre-substellar core candidate,  whose
properties seem to be consistent with a scaled-down version of low-mass stars.
\end{abstract}


\section{Introduction}\label{intro}

The formation mechanism of brown dwarfs (BDs) is currently a matter of vigorous
debate.  Since the typical thermal Jeans mass in molecular cloud cores
is around 1~\mo, a cloud should not fragment in cores of substellar
masses and the formation of brown dwarfs cannot be directly explained
as a scaled-down version of low-mass star formation.  A possible
scenario is that BDs form from fragmentation of turbulent
cores. In this scenario, turbulence produces particular density
enhancements, decreasing the Jeans mass and allowing the formation of
cloud cores of very low masses (Padoan \& Nordlund 2004). Another
possible scenario is the formation of BDs through
fragmentation of massive discs and subsequent ejection (Rice \et\
2003; Stamatellos \& Whitworth 2009), or ejection could also take
place in the very first stages of formation of a multiple system
(\eg\ Reipurth \& Clarke 2001; Umbreit \et\ 2005; Basu \& Vorobyov 2012), depriving the ejected objects from gaining further mass.
Other scenarios propose that an initially typical core is photoerosioned due to the nearby presence of a massive star
%impeding that the forming star gains a stellar amount of mass
(Whitworth \& Zinnecker 2004). 
%However, this mechanism cannot explain the formation of BDs in low-mass star-forming regions not containing any OB star.
%The most widely discussed formation scenarios include turbulent fragmentation (Padoan \& Nordlund 2004), ejection from multiple protostellar systems (Reipurth \& Clarke 2001, Bate et al. 2002), and photo-evaporation of massive pre-stellar cores (Whitworth \& Zinnecker 2004). 

An observational way to distinguish among these different scenarios is
the search and characterization of BDs in their very first
evolutionary stages, what we call here the proto-BD stage (following
the nomenclature of Pound \& Blitz 1993; 1995, and equivalent to Class
0/I stages of low-mass star formation).  If BDs form in a similar way
as their low-mass stellar counterparts, one would expect to find proto-BDs
surrounded by substantial discs and envelopes, with the envelopes
similar to the ones observed in the first stages of low-mass
protostars (\eg\ Andre, Ward-Thompson, \& Barsony 1993),
%Lee \et\ 2006)
which are strongly emitting in the millimetre/submillimetre range.  
In addition, proto-BDs should be found associated with outflow and ejection phenomena, 
whose parameters should be a scaled-down version of low-mass stars, and perhaps should even form planetary systems.
As a matter of fact, a significant number of BDs with ages of a few Myr
have been shown to have discs (\eg\ Luhman \et\ 1997; White \et\ 1999; Fern\'andez \& Comer\'on 2001; Barrado y Navascu\'es \& Mart{\'i}n
2003; Pascucci \et\ 2003; Barrado y Navascu\'es \et\ 2004, 2007; Natta \et\ 2004; Apai \et\ 2005; Mohanty \et\ 2005; Jayawardhana \et\ 2006a,b; Luhman \et\ 2005, 2007, 2008, 2010; Harvey \et\ 2010; Monin \et\ 2010; Joergens \et\ 2012; Riaz \et\ 2012b), 
but most of these discs were inferred from optical or near/mid-infrared data, 
being thus T-Tauri analogs or Class II sources.
%excluding thus the most embedded phases. 
%2006bApJ...648.1206 Jayawardhana, Ray; Coffey, Jaime; Scholz, Alexander; Brandeker, Alexis; van Kerkwijk, Marten H.	
%Accretion Disks around Young Stars: Lifetimes, Disk Locking, and Variability
%optical data
%
%2006aApJ...647L.167 Jayawardhana, Ray; Ivanov, Valentin D.	
%Spectroscopy of Young Planetary Mass Candidates with Disks
%optical data, Halpha
%
%2005MmSAI..76..295 Jayawardhana, Ray; Mohanty, Subhanjoy; Basri, Gibor	
%Accretion discs in the sub-stellar regime
%
%2003AJ....126.1515 Jayawardhana, Ray; Ardila, David R.; Stelzer, Beate; Haisch, Karl E., Jr.	
%A Disk Census for Young Brown Dwarfs
%
%2002ApJ...578L.141Jayawardhana, Ray; Mohanty, Subhanjoy; Basri, Gibor	
%Probing Disk Accretion in Young Brown Dwarfs

Different research groups have searched for proto-BDs in the
millimetre/submillimetre range.  One of the pioneering works was that
of Pound \& Blitz (1993; 1995), who find no clear-cut evidences of
proto-BDs. These works were followed up by Greaves \et\ (2003), who
survey an area of the Ophiucus star-forming region with the JCMT at
850~\mum, and find a dozen of possible planetary-mass isolated
objects. However, their submillimetre sources could be tracing either
truly proto-BDs or transient objects, and these objects remain to be confirmed through molecular line observations.
% which could be disrupted and never reach a collapsing state, as no mid/far-infrared point-like
%sources were associated with the submillimetre emission (except for Oph\,B-11, where outflow signposts were
%detected, and which might be considered as the first proto-BD candidate). 
%
Other candidates to be BDs in the Class 0/I phase are the so-called Very Low Luminosity Objects (VeLLOs), objects embedded in
dense cores with internal luminosities $<0.1\,L_\odot$ (Di Francesco et al. 2007; Dunham \et\ 2008), such as L1014-IRS and L1148-IRS, 
which are associated with dusty and dense envelopes, centimetre emission (L1014: Shirley \et\ 2007),
and CO\,(1--0) and  CO\,(2--1) outflows (Bourke \et\ 2005; Kauffmann \et\ 2011), all properties suggestive of accretion/ejection of matter.
However, for L1014-IRS and the other few VeLLOs studied in detail 
(\eg\ Andr{\'e}, Motte, \& Bacmann 1999; Young et al. 2004; Kauffmann et al. 2005, 2008, 2011; Lee et al. 2009; Dunham \et\ 2010) 
their true \emph{substellar} nature is not clear (eg., Lee 2007), nor from the bolometric luminosity ($\ga0.05$~\lo) neither from the mass of the envelope, which could increase the present mass well above the substellar limit once accreted on to the object (Kauffmann \et\ 2011).
%($M_\mathrm{env} \sim >70$~M$_\mathrm{jup}$).
%(Lee et al. 2009). 
%
On the other hand, the studies of Klein \et\ (2003),
%observed a sample of 19 young brown dwarfs using the JCMT and the IRAM\,30\.m Telescope. 
and Scholz \et\ (2006), who carry out surveys with the JCMT and/or the IRAM\,30\,m Telescope,
%at 1.3~mm in 20 young brown dwafs of Taurus using the IRAM\,30\,m Telescope. 
were addressed towards samples of relatively old brown dwarfs ($>1$~Myr), selected from optical/infrared surveys,
%Other individual detection of young BD candidates in the millimetre were made by Bouy \et\ (2008; 2M J044427). 
%Klein: ages of a few Myr; Scholz: ST later than M6, spectroscopically confirmed Taurus members), of a few Myr, 
%SHOWING POINT-LIKE EMISSION AND NO EXTENDED SOURCES IN THE MM RANGE?? WHICH IMPLIES A MORE ADVANCED EVOLUTIONARY STATE??
while the most embedded objects are expected at ages $\la0.5$~Myr.  
Finally, individual proto-BD candidates studied in detail in the millimetre/submillimetre range either are more evolved than Class I sources (\eg\ Phan-Bao \et\ 2008; Harvey \et\ 2012; Riaz \et\ 2012a), or do not have the mass well constrained and could still be stellar (\eg\ Scholz \et\ 2008).

%This situation has greatly improved with the Spitzer Space Telescope, which allows to select point-like embedded sources in the mid-infrared range, 
%and search for milllimeter/submillimetre emission associated with them, a new and very promising strategy (see, \eg\ Dunham \et\ 2008) to search for proto-BDs.  

In this paper we show the results of a millimetre/submillimetre study of a sample of 12 proto-BD
candidates selected in the B213-L1495 dark clouds of the Taurus-Auriga complex using Spitzer data. We show the results of the
observations at 1.2~mm toward this sample, as well as submillimetre
data for the two most promising candidates found at 1.2~mm.
One of the candidates was not previously reported, while the other candidate
was already reported in Barrado et al. (2009), and here we present subsequent observations aimed at further constraining its nature, which additionally reveal a nearby starless core of substellar mass, \ie\ a possible pre-substellar core.
%model which reproduces its SED and the radial intensity profile of its envelope at submillimetre wavelengths simultaneously.

%**NOTE: Star formation in the Taurus filament L1495: From Dense Cores to Stars; Schmalzl, Kainulainen, Quanz, Alves, Goodman, Henning, Launhardt, Pineda, Rom\'an-Z\'u\~niga 2011, ApJ, arXiv:1010.2755, estimate Av around 5 for the proto-BD position!!**

%To assess this ratio,  we downloaded the Spitzer data and selected the sources detected in the four IRAC bands, built a color-color diagram ([3.6]$-$[4.5] vs [5.8]$-$[8.0]) and classified the IRAC sources in the Class 0/I, Class I/II and Class II stages (following Allen \et\ 2004), and computed the ratio of Class 0/I to Class II sources. 
%For B213, we selected, among all the Class 0/I IRAC sources, those which seemed best candidates to be substellar objects (proto-BD candidates). For this, we built a magnitud-color diagram ([3.6] vs [3.6]$-$[4.5]) and compared the postiion of the IRAC sources with 1 Myr isochrones of the Dusty code (Baraffe \et\ 2003). In addition, we selected the IRAC sources with no 2MASS emission and with a steep spectral energy distribution (SED). Finally, we applied the rejection criteria of extragalactic contaminants from Gutermuth \et\ 2008). 

%Taurus BDs
%Monin, Guieu, Pinte et al. 2010, A\&A

\begin{figure}
\begin{center}
\begin{tabular}[b]{c}
    \epsfig{file=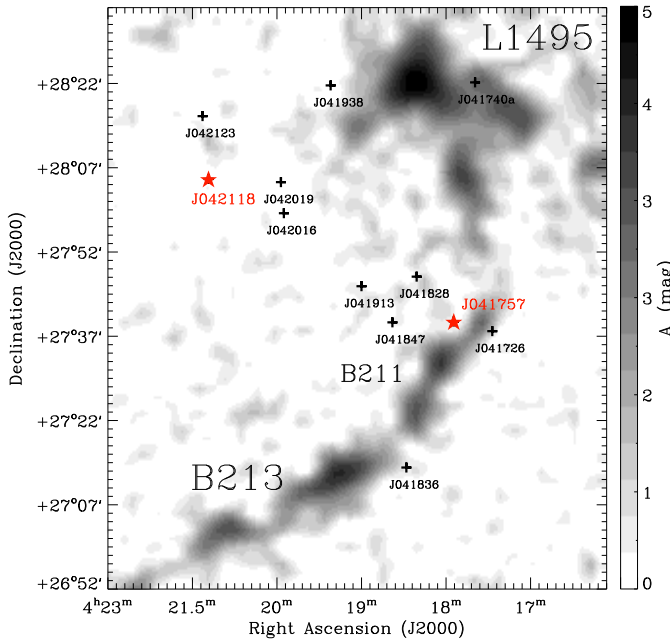, width=9cm,angle=0}\\
\end{tabular}
\caption{
Extinction map in 2MASS $J$-band derived with an adaptive star count method (Cambresy et al. 1997) in the B213-L1495 region ($45'$ radius). Members and candidate members from Kenyon \& Hartmann (1995) and references therein have been taken into account (removed from the star counting) when computing the $A_\mathrm{J}$ values. The resolution of the map varies from 30--50~arcsec, depending on the density of stars detected by 2MASS.  
%The Aj value in the proto-BD position is 0.35 approx, which corresponds to Av=1.0395. 
%Aqui esta la imagen de la extinci—n. Los valores son Aj (Aj/Av=2.97 --> I would say that this is typo: Aj/Av=1/2.97)
%(Obviamente la estructura circular es totalmente artificial; calcule los valores de forma radial y al recostruir la imagen... "pos claro", pasa lo que pasa :-)) 
Plus signs mark the positions of the selected Spitzer/IRAC sources observed with the IRAM\,30m Telescope, and the stars correspond to the sources additionally observed with the CSO. 
}
\label{fextinction}
\end{center}
\end{figure}

\section{Selection strategy and observations}\label{obs}

\subsection{Selection strategy}

We used IRAC and MIPS (Spitzer) data toward one of the youngest regions in Taurus, the B213-L1495
region. This region was selected from Froebrich (2005) list
of Class 0 sources, because it was one of the nearby ($<300$~pc) regions
with a higher ratio of Class 0/I to Class II IRAC sources (see Fig.~1 in Barrado \et\ 2009).
%and will be presented in more detail in
%Morales-Calder\'on \et\ (2010, in preparation). Here we present only a very brief summary.  
For B213-L1495, we downloaded the Spitzer archive
data and selected 12 IRAC sources which were classified as
Class 0/I according to IRAC color-color diagrams (Allen et al. 2004),
and substellar (with a magnitude below the 0.075~\mo\ cut in the
color-magnitude diagram, after comparing with evolutionary models, Barrado \et\ 2009). In addition, we rejected
those objects classified as extragalactic candidates following the
criteria of Harvey et al. (2006), Joergensen et al. (2006), and Bouy \et\ (2009).
%from a comparison between the c2d data (containing a fair
%amount of YSOs) and the data from SST SWIRE Legacy Project (essentially YSO-
%free sample). Figure 2b shows one of the two selection criteria used (IRAC at 8 ?m
%and MIPS at 24 ?m) to make the final target selection. The level of the criteria were
%chosen so that 95\% of the SWIRE sources are located in the area under the dashed lines.
Fig.~\ref{fextinction} shows the final sample overlaid on an extinction map of the region, elaborated for this work using the star count method on 2MASS data.
%From the extinction map of Schmalzl \et\ (2010) we can derive a visual extinction of $A_\mathrm{V}\sim5$... AT WHICH POSITION?

%Our selection strategy is similar to the strategy used by Dunham \et\ (2008). Both use an initial set of criteria to identify the candidate embedded objects from Spitzer data, and then carry out a subsequent study at longer wavelengths.  Dunham \et\ (2008) apply their criteria to the c2d Spitzer Legacy project and build a sample of 15 VeLLOs.  The main differences between our selection criteria and Dunham \et\ (2008) criteria is that they require a detection at 24 and 70~\mum, while we require a detection in the 4 IRAC bands.
%%(there is also a difference in the criteria adopted to reject extragalactic objects). 
%While Dunham's criteria allow to detect more embedded objects (as long as they do not require a detection in the 4 IRAC bands), our criteria allows us to build IRAC color-color and mag-color diagrams to select the Class 0/I substellar candidates and be more effective rejecting contaminants (because of Bouy+09).
%%J041757 and J042118 fulfill the selection criteria of Dunham \et\ (2008), except for criterion number 3 (detection
%%at 70 micron) and number 7 (J041757 is classified as galaxy under the
%%criterion of Harvey et al. 2007a, but see Barrado et al. 2009 for a
%%discussion on the possible extragalactic nature of J041757; J042118
%%SHOULD BE CHECKED). 

\subsection{IRAM\,30\,m ON-OFF at 1.2~mm and APEX at 870~\mum\ for the sample of 12 proto-BD candidates \label{somambo}}

We observed with the IRAM\,30\,m\footnote{The IRAM\,30m Telescope is operated by the Institut de Radioastronomie
Millimetrique, which is supported by INSU/CNRS (France), MPG (Germany),
and IGN (Spain).} single-dish Telescope (Pico Veleta,
Spain) the 1.2~mm continuum emission of the 12 selected BD candidates using
the 117-channel MAMBO-II bolometer array.
%which has a for which every pixel has a
The Half Power Beam Width (HPBW) of the telescope at 1.2~mm  is 11~arcsec. 
Observations were carried out in 2007 May, October, November, and 2008 November 14, and the opacities ranged between 0.30 and
0.40. Each source was centred in the most sensitive pixel of the
array (pixel 20). Typical pointing corrections were $\la5$~arcsec,
providing a positional accuracy of $\sim3$~arcsec. The background
subtraction was performed by using the ON-OFF observing mode, with a
wobbler throw of 32~arcsec. Every source was observed for at least one
scan of 20 minutes with an integration time per subscan of 60\,s.
The data were reduced using the MOPSIC pipeline provided by IRAM.
In order to test the consistency of the data we measured for each
source the flux density for each scan separately, and obtained values
that were consistent within the uncertainties. In Table~\ref{tobs} we summarize the observations presented in this work, and in
Table~\ref{tmambo} we give the list of the 12 selected Spitzer/IRAC sources\footnote{Throughout this paper, we will refer to each source of Table~\ref{tmambo} designated by `SSTB213\,J04mmss.ss+ddmmss.s' as `J04mmss'.}, total integration time, opacity at 250~GHz, and the measured flux density at 1.2~mm. The final rms noise at 1.2~mm was
typically 0.7--1.5 m\jpb, depending on the integration time. None of the 12 Spitzer/IRAC sources was detected above the $4\sigma$ level, corresponding to envelope/disk masses $\la3$~\mj\ (for a dust temperature of 15~K and a dust opacity at 1.2~mm of 0.009 cm$^2$\,g$^{-1}$, Ossenkopf \& Henning 1994, Table~\ref{tmambo}). However, for two of the observed Spitzer/IRAC sources we measured fluxes above 2$\sigma$, namely J041757 and J042118, which were subsequently observed with the CSO.

In addition, we used public archived data from LABOCA for the Atacama Pathfinder Experiment (APEX) antenna\footnote{APEX is a collaboration between the Max-Planck-Institut fuer Radioastronomie, the European Southern Observatory, and the Onsala Space Observatory.}. Observations at 870~\mum\ were carried out  in 2009 July under the project E-083.C-0453A.2009 (P.I. M. Tafalla). The region mapped corresponds to Taurus B213 and B211 filaments, and was observed during a total time of 14~hours, under good weather conditions (precipitable water vapor (pwv) ranged between 0.25 and 1.8~mm). Data reduction was performed using BoA and miniCRUSH software packages following standard procedures. The LABOCA image covers 5 of the 12 Spitzer/IRAC sources of our sample, and none of the 5 sources were detected (see Table~\ref{tmambo} for upper limits).

%LABOCA
%12, 20,22,28 y 29 de Julio del 2009, ~14 horas de tiempo total incluyendo overheads. La pwv vario entre 0.25 y 1.8 mm 

%AKARI: looking for possible sources falling within the wobbler throw of IRAM30m/MAMBO observations:
%In order to assess the possibility that the wobbler throw was merging emission from other nearby sources, we checked the infrared AKARI images both in the FIS and IRC instruments (thus covering wavelengths from 9 to 160~$\mu$m), and found no AKARI point sources within 2 arcmin of radius for any of the targets except for J041740, where a source at 90, 140 and 160~\mum\ was found, J0417426+282442. This AKARI source (with no associated object in the Simbad Database\footnote{http://simbad.u-strasbg.fr/simbad/}) is $\sim31$~arcsec to the west and $\sim26$~arcsec to the north, with respect to the IRAC target source. The brightest flux of this source, 3.6 Jy, is found at 140~\mum. Assuming blackbody emission with a spectral index of 3--4, this would correspond to a flux density at 1.2 mm of 0.6--6 mJy. Hence, J0417426+282442 will not be taken into account to compute statistics.  
%{\it We should also check how isolated are our targets by checking the Spitzer images (all filters) if there is any infrared source within an aperture of $64''$ (twice the wobbler throw) which could be contaminating. MARIA, COULD YOU PROVIDE US WITH THE SPITZER IMAGES?}

\begin{table*}
\caption{Summary of observations and catalogs used in this work}
\begin{center}
{\small
\begin{tabular}{lcccccc}
\noalign{\smallskip}
\hline\noalign{\smallskip}
&Wavelength
&Observing
&Sources
&Date
&HPBW\supa
&P. A.\supa
\\
Telescope/Instrument
&(\mum)
&mode
&observed
&(yy/mm/dd)
&(arcsec)
&(\degr)\\
\noalign{\smallskip}
\hline\noalign{\smallskip}
CFHT/{\scriptsize MegaPrime-MegaCam}&0.75,0.90&photometry&J041757&04/12/05\supb	&$\sim1$			&$-$\\
APO/SDSS		&0.89		&photometry	&J042118			&$>2004$			&$\sim1$			&$-$\\
CAHA/Omega2000	&1.25,1.65,2.17&photometry	&J041757			&07/11/04\supb		&$\sim1$			&$-$\\
UKIDSS/WFCAM	&1.25,1.63,2.20&photometry	&J042118			&07/09/28			&$\sim1$			&$-$\\
Spitzer/IRAC		&3.6,4.5,5.8,8.0&photometry	&Table~2			&05/02/21\supb		&1.4--1.9			&$-$\\
WISE/SC			&12,22		&photometry	&J041757,J042118	&2010			&6.5--12			&$-$\\
Spitzer/MIPS		&24,70		&photometry	&Table~2			&05/02/28\supb		&6--18			&$-$\\
CSO/SHARC-II     	&350		&cont. imaging	&J041757,J042118	&08/01/28			&10				&$-$\\
APEX/LABOCA	&870		&cont. imaging	&5 sources Table~2	&09/07			&18				&$-$\\
IRAM\,30m/MAMBO-II&1200		&ON-OFF 	&Table~2  		&07/05,07/11		&11				&$-$\\
IRAM\,30m/EMIR	&2700,\tco(1--0)&line imaging	&J041757			&09/10/28			&22				&$-$\\
IRAM\,30m/EMIR	&2600,\co(1--0)&line imaging	&J041757			&09/10/28			&22				&$-$\\
IRAM\,30m/EMIR	&3200,\nth(1--0)&freq. switch.	&J041757			&10/08/13			&27				&$-$\\
VLA				&36000		&cont. imag.,D\supc	&J041757		&08/09/04			&$10 \times 9$\phb	&$-68$\degr \\
VLA				&36000		&cont. imag.,B\supc	&J041757		&09/02/25			&$1.7\times1.6$ 	&$-89$\degr \\
VLA				&60000		&cont. imag.,D\supc	&J041757		&08/09/04			&$17\times16$ 	&$+77$\degr \\
VLA				&60000		&cont. imag.,B\supc	&J041757		&09/02/25			&$2.4\times1.8$ 	&$+61$\degr \\
\hline
\end{tabular}
\begin{list}{}{}
\item[$^\mathrm{a}$] Half Power Beam Width (HPBW) and Position Angle (P. A.) of the beam.
\item[$^\mathrm{b}$] Some data already published in Barrado \et\ (2009).
\item[$^\mathrm{c}$] `B' or `D' stands for the array configuration. 
\end{list}
}
\end{center}
\label{tobs}
\end{table*}

\begin{table*}
\caption{IRAM\,30m/MAMBO-II at 1.2~mm and Archive APEX/LABOCA at 870~\mum\ observations for the sample of proto-BD candidates}
\begin{center}
\begin{tabular}{lccccccc}
\noalign{\smallskip}
\hline\noalign{\smallskip}
&\multicolumn{2}{c}{Position}
&Int. time\supa
&$S_\nu^\mathrm{1.2mm}$\,\supb
&$M_\mathrm{env}$\,\supc
&$S_\nu^\mathrm{870\mu m}$\,\supd
\\
\cline{2-3}
Source
&$\alpha (\rm J2000)$
&$\delta (\rm J2000)$
&(min)
%&$\tau_{250~\mathrm{GHz}}$
&(mJy)
&(\mj)
&(mJy)
\\
\noalign{\smallskip}
\hline\noalign{\smallskip}
SSTB213\,J041726.38+273920.0  	&04 17 26.38   	&+27 39 20.0	&40		&$<4.2$\phda	&$<2.7$	&$<\phda84$	\\
SSTB213\,J041740.32+282415.5	&04 17 40.32	&+28 24 15.5	&40	 	&$<4.2$\supe	&$<2.7$	&$-$		\\
SSTB213\,J041757.77+274105.0	&04 17 57.77	&+27 41 05.0	&100  	&$<2.9$\supb	&$<1.8$	&$<\phda84$	\\
SSTB213\,J041828.08+274910.9  	&04 18 28.08  	&+27 49 10.9	&20	  	&$<5.4$\phda	&$<3.4$	&$<480$	\\
%J041828 detected with CAHA Omega2000!!
SSTB213\,J041836.33+271442.2  	&04 18 36.33  	&+27 14 42.2	&20	  	&$<7.2$\phda	& $<4.6$	&$<\phda96$	\\
SSTB213\,J041847.84+274055.3  	&04 18 47.84  	&+27 40 55.3	&20	  	&$<5.8$\phda	& $<3.7$	&$<\phda88$	\\
%J041847 detected with CAHA Omega2000!!
SSTB213\,J041913.10+274726.0  	&04 19 13.10     &+27 47 26.0	&20	  	&$<5.9$\phda	&$<3.7$	&$-$		\\
SSTB213\,J041938.77+282340.7  	&04 19 38.77  	&+28 23 40.7	&20	  	&$<5.0$\phda	&$<3.2$	&$-$		\\
SSTB213\,J042016.70+280033.7  	&04 20 16.70  	&+28 00 33.7	&20	  	&$<5.7$\phda	&$<3.6$	&$-$		\\
SSTB213\,J042019.20+280610.3  	&04 20 19.20  	&+28 06 10.3	&20	  	&$<5.4$\phda	&$<3.4$	&$-$		\\
SSTB213\,J042118.43+280640.8 	&04 21 18.43   	&+28 06 40.8	&40	 	&$<4.0$\supb	&$<2.5$	&$-$		\\
SSTB213\,J042123.70+281800.6  	&04 21 23.70  	&+28 18 00.6	&20	  	&$<5.6$\phda	&$<3.5$	&$-$		\\ 
\hline
\end{tabular}
\begin{list}{}{}
\item[$^\mathrm{a}$] Total integration time for IRAM\,30m/MAMBO-II ON-OFF observations.
\item[$^\mathrm{b}$] Upper limits correspond to $4\sigma$. For the case of J041757 and J042118, the weighted mean of the different subscans was $1.89\pm0.72$ and $2.15\pm1.00$~mJy respectively, both above $2\sigma$, while for the other sources the weighted mean was well below $2\sigma$.
\item[$^\mathrm{c}$] Masses of disc/envelope estimated from the 1.2~mm continuum emission, assuming the emission is optically thin, a dust temperature of 15~K, and a dust opacity at 1.2~mm of 0.009 cm$^2$\,g$^{-1}$ (Ossenkopf \& Henning 1994).
\item[$^\mathrm{d}$] Upper limits correspond to $4\sigma$, and are measured in the Archive APEX/LABOCA image of the B213/B211 filament (see \S~\ref{somambo}).
\item[$^\mathrm{e}$] No reliable flux due to the presence of a submillimetre source (J0417426+282442, detected by AKARI with a flux of 3.6 Jy at 140~\mum) within the wobbler throw. We estimated that up to 6~mJy could have been subtracted to this source. This is the only source of our sample with an AKARI submillimetre source within the wobbler throw of the IRAM\,30m/MAMBO observations.
\end{list}
\end{center}
\label{tmambo}
\end{table*}

\subsection{CSO at 350~\mum\ for J042118 and J041757 \label{socso}}

The Caltech Submillimetre Observatory (CSO) observations toward J042118 and J041757 were carried out on 2008 January 28, using the SHARC-II bolometer array at
350~$\mu$m. The HPBW of the CSO at this wavelength is 8.6~arcsec, which
was convolved to a final beam of 10.4~arcsec. Typical opacities at
225~GHz were around 0.04.  The observations were carried out in
imaging mode using a lissajous pattern of 20~arcsec of amplitude and
14.142 seconds of period, resulting in a final image size of
$3.3\times3.2$~arcmin$^2$. In this observing mode, the telescope modulates the
X and Y position with a different sine wave for each axis (starting
from the centre of the map). As a consequence, the rms is lower in the centre of the image than in the edges.
%is not uniform across the image, being lower at the centre of the image. 
%In this type of observations, the sky noise at those frequencies cannot be separated from the signal?????. 
Each scan was integrated for 10 minutes, and the total observing time per source was around 1--1.5 hours.  The absolute flux uncertainty was estimated from a measurement of the flux of HL\,Tau, and was $\la10$\%.  The uncertainty in position was $\sim3$~arcsec. 
The reduction of the SHARC-II data was carried out with the CRUSH software (Kov\'{a}cs 2008), version 1.62-4. For both sources we assumed that the emission is compact ($<30$~arcsec) and faint  ($<1$~Jy)\footnote{In Barrado \et\ (2009) we show the CSO emission for J041757 assuming that the source is extended, i.e., allowing to recover emission from structures $>30$~arcsec. For this case the source associated with J041757 is $\sim30$~arcsec, and has a small extension to the north-east. However, the `compact' option used in this work, although yielding more negative features, recovers a structure which is more consistent with independent observations of the region 
(as shown in this paper). For the J042118 region, the assumption of  `extended' emission (forcing the software to recover structures $<60$~arcsec to avoid sky gradients) yields very similar results to the assumption of `compact' emission.}. The rms noises achieved at the map centres are 9~mJy for J042118 and 6~mJy for J041757.

\subsection{VLA observations at 6 and 3.6~cm for J041757 \label{sovla}}

We obtained radio continuum data at 6~cm (4.86 GHz) and  3.6 cm (8.44 GHz) towards the source J041757 using the Very Large Array
(VLA) of the National Radio Astronomy Observatory (NRAO)\footnote{The National Radio Astronomy Observatory is a facility of the National Science Foundation operated under cooperative agreement by Associated Universities, Inc.}.  Observations were performed on 2008 September 4 (project AP566), and on 2009 February 25 (project AD593), with the array in D and B configuration respectively. We used the standard VLA continuum mode (4IF, 50 MHz per IF), and both right and left polarizations were processed. The phase centre of the observations was R.A.(J2000.0)=04$^{h}$17$^{m}$57.8$^{s}$, Dec(J2000.0)=$+$27$^{\rm o}$41$'$05.0$''$ in both epochs.
For the first set of data (D configuration of the array) and for each
frequency, we used $\simeq$ 40 minutes on source, and for the second set (B configuration) we integrated $\simeq$ 1.5 hour on source at each frequency. See Table~\ref{tvlaobs} for further details on the observations.
%The primary calibrator was 3C147 (adopted flux density of 7.9 Jy at 6 cm, and 4.7 Jy at 3.6 cm) and the phase calibrator was J0431+206, for which a bootstrapped flux of $2.34\pm0.03$ at 6~cm and $1.590\pm0.005$~Jy at 3.6~cm was determined.  For the second set of data (B configuration), we integrated $\simeq$ 1.5 hour on source at each frequency. The quasar 3C138 was selected as the flux calibrator for these second observations, for which we adopted a flux density of 3.6 Jy and 2.4 Jy respectively for 6 and 3.6 cm wavelengths. The bootstrapped flux of the phase calibrator J0431+206 in B configuration was $2.63\pm0.05$ at 6~cm and $1.06\pm0.04$~Jy at 3.6~cm. The synthesized beams for each wavelength at each configuration are listed in Table~\ref{tobs}. 

Both sets of data were obtained during the Extended Very Large Array
(EVLA) transition mode. Calibration and data reduction were
performed using the Astronomical Image processing System (AIPS, developed by NRAO), and following
NRAO official recommendations for the EVLA transition mode\footnote{During the transition to the EVLA, the 50~MHz
continuum mode was affected by high closure errors on EVLA-VLA
baselines (affecting strongly the 3.6~cm receiver; for this reason, Barrado \et\ (2009) report only the flux density at 6~cm).
This problem is caused by non-matched bandpass shapes of
different type of antennas. In this work, we tried to fix this problem by observing a strong
source with known structure and determining baseline-based closure
corrections using the AIPS task BLCAL.}.
As a result, data in D configuration at both 6 and 3.6~cm were successfully calibrated. 
As for the second dataset (B configuration), anomalous data affecting the flux calibrator were flagged at 6~cm\footnote{For the B configuration data, anomalous jumps in the amplitudes and phases for certain EVLA antennas, specially during the end of the target and flux calibrator observation, prevented us from a reliable flux calibration. These anomalous data were flagged only for the 6~cm case, because the 3.6~cm observations were too short. However, flagging these data with anomalous jumps improved the signal-to-noise, while the morphology of the object was found unaffected, giving us confidence on the morphology of the source seen in the B configuration data at both 6 and 3.6~cm.}.
Thus, we used data in B configuration just to illustrate the morphology of the emission
at higher angular resolution, and for flux measurement purposes we used only the first dataset (D configuration).
For the D configuration data, we produced radio continuum maps with natural weighting to optimize the sensitivity of the images, achieving rms noises of 
$\simeq$0.025 and $\simeq$0.055 m\jpb\ at 6~cm and 3.6~cm, respectively.  For the B configuration data, we produced maps with natural weighting and using a 
taper in the $uv$-data of 150~k$\lambda$ at 6~cm and 120~k$\lambda$ at 3.6~cm. For the B configuration data we refrain from giving the rms noises because of anomalous data for the flux calibrator.

\begin{table}
\caption{Parameters of the VLA radio continuum observations towards J041757}
\begin{center}
{\small
\begin{tabular}{ccccccc}
\noalign{\smallskip}
\hline\noalign{\smallskip}
&$\lambda$
&Flux
&Flux
&Phase
&Flux\,\supa
\\
Config.
&(cm)
&calib.
&(Jy)
&calib.
&(Jy)
\\ 
\noalign{\smallskip}
\hline\noalign{\smallskip}
D	&6     	&3C147  &7.9       &J0431+206	&$2.34\pm0.03$\\
D	&3.6     	&3C147  &4.7       &J0431+206	&$1.590\pm0.005$\\
\hline
B	&6     	&3C138  &3.6       &J0431+206	&$2.63\pm0.05$\\
B	&3.6     	&3C138  &2.4       &J0431+206	&$1.06\pm0.04$\\
\hline
\end{tabular}
\begin{list}{}{}
\item[\supa] Bootstrapped flux.
\end{list}
}
\end{center}
\label{tvlaobs}
\end{table}

\subsection{IRAM\,30m CO, $^{13}$CO and N$_2$H$+$ emission for J041757}

J041757 was observed in the CO\,(1--0) (115.271204~GHz), and \tco(1--0) (110.201370~GHz)
%and CO\,(2--1)
transitions with the IRAM\,30\,m Telescope using the EMIR (Carter et al. 2012) heterodyne receivers.
%and the 18 element heterodyne receiver array HERA (Schuster et al. 2004). 
The observations were carried out on 2009 October 28 and 29,
with system temperatures of 200--400~K ($\tau \sim0.16$--0.35).
%, and 400--600~K ($\tau \sim0.24$) for HERA.
%around 400--500~K for HERA1 and 500--600~K for HERA2, and
We conducted on-the-fly maps in position-switching mode, scanning the region in R.\,A., and with steps of 4~arcsec in Dec., covering a size of
$1.6\times1.6$~arcmin$^2$ centred on J041757.
%, which engulfs well the submillimetre extended emission previously observed with the CSO. 
%Steps in DEC were of 4~arcsec, which corresponds to Nyquist sampling for the 11~arcsec beam of the telescope at
%the CO\,(2-1) frequency.
For each of the CO\,(1--0), and $^{13}$CO(1--0) transitions we used two
units of the VESPA (VErsatile SPectrometer Array) backend, one in each polarization, of 40~kHz (0.11~\kms) of spectral resolution and a 
bandwidth of 40~MHz ($\sim100$~\kms).  Simultaneously to the VESPA backend, we also used the WILMA (WIdeband Line Multiple Autocorrelator) backend, providing a bandwidth of 9000~\kms\ with $\sim5$~\kms\ of spectral resolution at the frequency of the CO\,(1--0) transition.
%As for the CO\,(2--1) transition, we used two units of VESPA of 80~kHz (0.10~\kms) of
%spectral resolution and 80 MHz ($\sim100$~\kms) of bandwidth. 
%For CO\,(1--0) we used the E090 receiver tuned at 115.271204 in Horizontal and Vertical
%polarizations. We simultaneously observed the H2CO line at 145.602952
%using the E150 receiver also in Horizontal and Vertical
%polarizations. However, we had to observe with WILMA as backend, which
%provided a bad spectral resolution. Regarding the
%$^{13}$CO(1--0), we observed it using the E090 receiver tuned at
%110.201370 with the same VESPA units as for the CO. We simultaneously
%observed the HCO+(3-2) at 267.557625 using the E230 receiver. However,
%the poor weather conditions and the fact that the line falls at the
%edge of the band did not allow to properly study the HCO+ emission
%(Tsys was around 2300 K).
The OFF position used,  at ($-7600,-287$)~arcsec from the map centre, was selected as free of CO from a survey in Taurus (Goldsmith \et\ 2008), and was proved to be free of emission.
%in CO\,(2--1) 
Calibration following the standard procedures was carried out during observing time and using the MIRA
package, and typical calibration errors are within 15--20\%. 
To make the final map, spectra were averaged every 8~arcsec in both RA and DEC.
% for the CO\,(1--0) and $^{13}$CO\,(1--0) maps., and every 4~arcsec for the CO\,(2--1) map.
Typical rms noise of spectra of each point of the final map were around 0.24, and 0.12~K, 
%and 0.5 K (Ta*), 
for CO\,(1--0), and $^{13}$CO\,(1--0), respectively,
%and CO\,(2--1), 
in main beam brightness temperature scale.
The main beam efficiency at the CO\,(1--0) line frequency is 0.82 ($F_\mathrm{eff}=$94.01\%, $B_\mathrm{eff}=$77.53\%) and at the \tco(1--0) frequency is 0.83
($F_\mathrm{eff}=$94.18\%, $B_\mathrm{eff}$=78.13\%).

Subsequent observations of the dense gas tracer N$_2$H$^+$(1--0) (at 93.173764~GHz) were carried out on 2010 August 13 under reasonable summer weather conditions (pwv $\sim4$~mm; system temperature $\sim 130$~K; opacity $\sim0.4$), in frequency switching mode, and using VESPA with 40~MHz of bandwidth and 0.02~MHz of spectral resolution. We integrated at different positions in and around J041757, separated $\sim20$~arcsec and covering offsets with respect to J041757 from (40,40)~arcsec to ($-20,-20$)~arcsec, reaching rms noises in each individual spectra of $\sim0.015$~K (main beam brightness temperature). The main beam efficiency at the \nth(1--0) frequency is 0.85 ($F_\mathrm{eff}=$94.76\%, $B_\mathrm{eff}=$80.15\%).
%C18O(1--0) --> wilma1, wilma2 --> bad spectral resolution! 
%N2H$^+$(1--0) --> vespa1, vespa2
All data reduction was done using the GILDAS software package supported at IRAM.

%\subsection{EVLA and NASA\,70m NH3(1,1) emission from the strongest CSO object}

\section{Results}\label{res}

\subsection{J042118 \label{srj042}}

\begin{figure}
\begin{center}
\begin{tabular}[b]{c}
   \epsfig{file=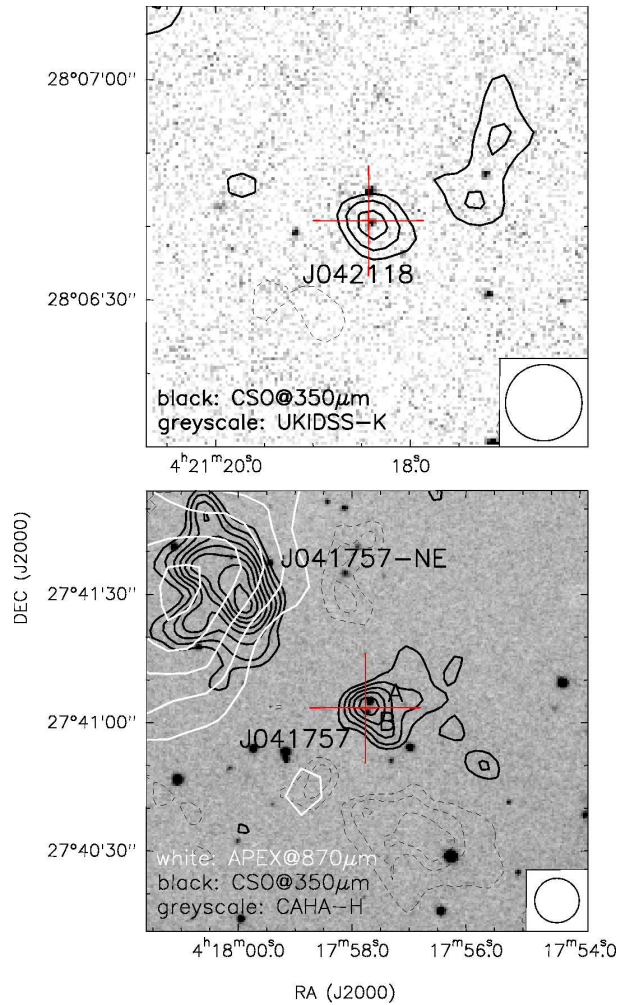, width=8cm,angle=0}\\
\end{tabular}
\caption{CSO 350~\mum\ signal-to-noise contours superimposed on a NIR image.
{\bf Top:} J042118. Contours are $-4$, $-3$ (dashed), 3, 4, and 5 (solid) times the rms noise of the map (noise is variable across the map, from 9 (centre) to 11 m\jpb\ (wedges)). 
Greyscale is the UKIDSS $K$-band image.
{\bf Bottom:} J041757. Black contours correspond to the CSO 350~\mum\ emission and are $-5$, $-4$, $-3$ (dashed), 3, 4, 5, 6, 7, 8, 9 and 10 (solid) times the rms noise of the map (noise is variable across the map, from 6 (centre) to 9 m\jpb\ (wedges)). White contours correspond to the APEX 870~\mum\ emission and are 1, 2, 3, 4, and 5 times 19~m\jpb. The APEX beam at 870~\mum\ is 18~arcsec.
Greyscale is CAHA/Omega2000 $H$-band image (see Barrado \et\ 2009 for details on the observations), showing the two NIR sources, A and B, found near the IRAC source.
In both panels, the CSO beam, shown in the bottom-right corner, is 10~arcsec, and the crosses indicate the position of the Spitzer/IRAC source.
}   
\label{fcso}
\end{center}
\end{figure}

Although the 1.2~mm emission for J042118 was very marginal (2.2$\sigma$, Table~\ref{tmambo}), the CSO observations at 350~\mum\ reveal one clear source with a flux level of $5\sigma \sim 46$~mJy (Fig.~\ref{fcso}-top). The position of the submillimetre source, which is unresolved, is coincident with the position of the Spitzer/IRAC and UKIDSS source, and there is marginal extended and faint emission at 350~\mum\ about 10~arcsec to the west.
In Table~\ref{tcso} we give the position, size, flux density and estimated mass of the submillimetre source detected in J042118. The mass of the disc/envelope obtained for J042118 is in the range 0.3--3~\mj, assuming a distance of 140~pc, a dust temperature in the range 10--20~K, and a dust opacity law at 350~$\mu$m of 0.101~cm$^2$g$^{-1}$ (Ossenkopf \& Henning 1994). Such a mass is at least 3 orders of magnitude larger than the disc masses estimated from Herschel observations for a small sample of young BDs (Harvey \et\ 2012).

\begin{table*}
\caption{Parameters of the sources detected with CSO/SHARC-II at 350~$\mu$m}
\begin{center}
{\small
\begin{tabular}{lccccccccc}
\noalign{\smallskip}
\hline\noalign{\smallskip}
& rms$^\mathrm{a}$
&\multicolumn{2}{c}{Position$^\mathrm{b}$}
&Deconvolved
&Deconvolved
&Deconvolved
&$I_\mathrm{\nu}^\mathrm{peak}$~$^\mathrm{c}$
&$S_\mathrm{\nu}^\mathrm{c}$
&Mass$^\mathrm{d}$
\\
\cline{3-4}
Source
&(mJy)
&$\alpha (\rm J2000)$
&$\delta (\rm J2000)$
&ang.size (arcsec)
&P. A. (\degr)
&size (AU)
&(m\jpb)
&(mJy)
&(\mj)\\
\noalign{\smallskip}
\hline\noalign{\smallskip}
%J042118-ext   & 7.3 &04:21:18.472 &28:06:40.900 &$8.3\times 6.8$  &$1200 \times 950$          &$26\pm7$      &26 &0.1--0.8\\ 
J042118	 	&8.9	&04:21:18.38 	&28:06:40.1 &$10.5\times8.7$\phe 	&\phb58.8 &$1500 \times 1200$	&$46\pm9$      	&46 &0.3--3\\
%J041757-ext   & 6.4 &04:17:57.178 &27:41:08.834 &$35.8\times 29.1$ &$4900 \times 4100$    &$103\pm6$   &1080 &4.5--32\\
J041757		&6.1	&04:17:57.67 	&27:41:05.8 &$19.5\times 12.7$ 	&$-67.3$	&$2700 \times 1800$ 	&$37\pm6$     	&165 &1--10\\
J041757-NE	&8.6	&04:18:00.30	&27:41:36.3 &$29.2\times 15.4$	&\phb29.7	&$4100 \times 2100$	&$84\pm9$	&356 &2--21\\
\hline
\end{tabular}
\begin{list}{}{}
\item[$^\mathrm{a}$] Rms noise at the position of the source. Total integration time was around 1h 30min for J041757 and around 1h for J042118. 
\item[$^\mathrm{b}$] Positions corresponding to the intensity peak.
\item[$^\mathrm{c}$] Peak intensity and flux density derived by computing statistics within the aperture corresponding to the 3$\sigma$ level of the source. 
\item[$^\mathrm{d}$] Masses derived assuming a dust temperature of 10--20~K, and a dust mass opacity coefficient from Ossenkopf \& Henning
(1994, see main text). The uncertainty in the masses due to the opacity law is estimated to be a factor of 2.
%\item[$^\mathrm{f}$]  Intensity and flux density derived from a Gaussian fit (MM6).
\end{list}
}
\end{center}
\label{tcso}
\end{table*}

\subsection{J041757 \label{srj041}}

\subsubsection{CSO 350~\mum\ and APEX 870~\mum\ continuum emission \label{srj041cso}}

J041757, with a 2.6$\sigma$ flux at 1.2~mm (Table~\ref{tmambo}), is well detected at 350~\mum, and presents a complex emission, with one partially resolved source ($\sim16$~arcsec of size) associated with the selected IRAC source J041757, at the centre of the field of view, and one stronger and more extended source ($\sim20$~arcsec) to the north-east (hereafter, J041757-NE), which is elongated in the southwest-northeast direction (Fig.~\ref{fcso}-bottom and Table~\ref{tcso}).
%Using the extended assumption, the source is detected at 18$\sigma$, with a peak intensity of
%$103\pm6$~m\jpb. The emission is around 30~arcsec in size and has some substructure, much larger than the J042118.  
The flux density of J041757 at 350~$\mu$m is 165~mJy, yielding a mass of the envelope of 1--10~\mj, (for a dust temperature of 10--20~K). Assuming a mean molecular weight (of H$_2$ molecule) of 2.8~m$_\mathrm{H}$, and a size (diameter) of 2200 AU, we obtain a density of $\la2.1\times10^5$~\cmt, a H$_2$ column density of $\la4.6\times10^{21}$~cm$^{-2}$, and $A_\mathrm{v}\la4.9$~mag (following Frerking \et\ 1982).  Similarly, for J041757-NE we estimate a mass of 2--20~\mj, a density of $\la2.4\times10^5$~\cmt, a H$_2$ column density of $\la6.4\times10^{21}$~cm$^{-2}$, and $A_\mathrm{v}\la6.7$~mag. These $A_\mathrm{v}$ values are in perfect agreement with the values derived by Schmalzl \et\ (2010) for this region.
%Out of the total visual extinction estimated from the 350~\mum\ emission for each source, about $\sim1$~mag corresponds to interstellar extinction in V (in the direction of J041757, the star count map indicates $A_\mathrm{J}\sim0.3$~mag, and $A_\mathrm{J}/A_\mathrm{V}\sim0.31$, see Fig.~\ref{fextinction}).
It is interesting to note that J041757-NE is also recovered in the archived APEX/LABOCA 870~\mum\ image (see Fig.~\ref{fcso}-bottom), for which we measured a peak intensity of 100~m\jpb, and a flux density (within the 2$\sigma$ contour) of 420~mJy. This corresponds to a mass of 75--160~\mj\ (assuming a dust temperature of 10--15~K,
%(Flagey \et\ 2009)
and a dust opacity law at 870~$\mu$m of 0.018 cm$^2$g$^{-1}$, Ossenkopf \& Henning 1994). The mass of the J041757-NE source as estimated from the APEX/LABOCA image is about a factor of 8 higher than the mass estimated from the CSO because structures $>30$~arcsec were filtered out during the reduction process of the CSO data (\S~\ref{socso}). Thus, the total mass of J041757-NE is $\sim100$~\mj, from which $\sim10$~\mj\ are found in a compact structure.

%***DEFINITIVE RMS NOISE OF THE J041757 CSO IMAGE (SEE ABOVE): 0.014 mJy -> BUT I MEASURED IT IN THE SUBIMAGE PROVIDED BY CRUSH
%Adopting a 10\% of uncertainty in the flux scale of the 103 mJy flux,
%the total uncertainty in the 103 mJy value is 17 mJy.***

\subsubsection{VLA 3.6 and 6~cm continuum emission \label{srvla}}

\begin{figure}
\begin{center}
\begin{tabular}[b]{c}
    \epsfig{file=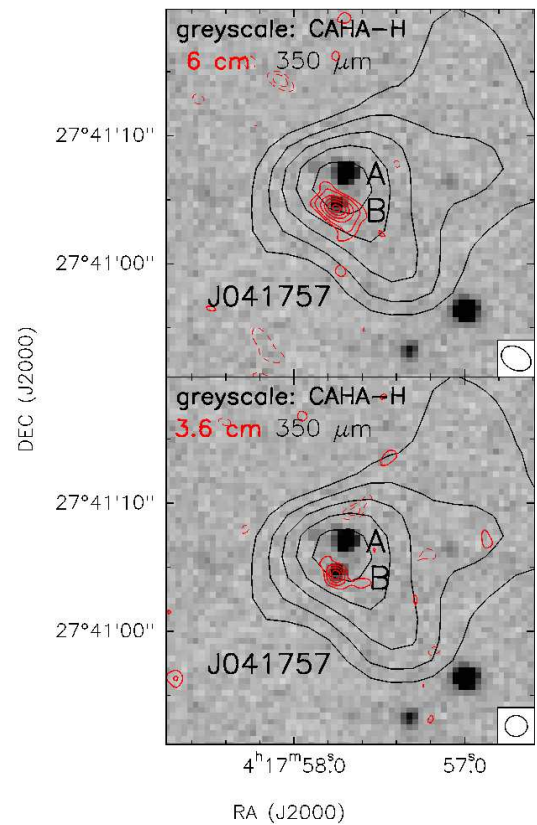, width=6.9cm, angle=0}\\
\end{tabular}
\caption{
High angular resolution VLA radio continuum emission (red contours) towards J041757, obtained with the array  in B configuration, superimposed on the Omega2000 $H$-band image (greyscale) and the CSO 350~\mum\ image (black contours).  
{\bf Top:} 6~cm emission. Red contours are $-4$, $-3$ (dashed), 3, 4, 6, 8, 10, and 12 (solid) times the rms of the map (rms not given due to technical problems with the flux density scale during the B-configuration observations, see \S~\ref{sovla}).
%, $3.7\times10^{-5}$~\jpb. 
The synthesized beam is shown in the bottom-right corner, and is $2.45\times1.77$~arcsec$^2$, with PA=61\degr. 
{\bf Bottom:} 3.6~cm emission. Red contours are $-4$, $-3$ (dashed), 3, 4, 5, and 6 (solid) times the rms of the map.
%, $1.4\times10^{-5}$~\jpb. 
The synthesized beam is shown in the bottom-right corner, and is $1.74\times1.64$~arcsec$^2$, with PA=$-89$\degr. 
In both panels, the greyscale and black contours are the same as in Fig.~\ref{fcso}-bottom.
% corresponds to the CAHA K-band image reported in Barrado et al. (2009, labels are the NIR sources A and B) and the grey contours correspond to the Herschel emission at 160~\mum\ already shown in Fig.~\ref{fpacscso}.
}   
\label{fvla}
\end{center}
\end{figure}

The VLA centimetre emission around J041757 reveals one single source at both 3.6 and 6 cm wavelengths at the position of J041757, with no significant emission at the position of J041757-NE.  
Using D-configuration only, the emission is unresolved at both frequencies, and it is
detected at $\simeq$7$\sigma$ at 6~cm, and marginally
detected at $\simeq$4$\sigma$ at 3.6~cm, with a flux density at both wavelengths of $\sim0.2$~mJy (Table~\ref{tvla}).  
The spectral index in the range of 6 to 3.6~cm is $\alpha_{6-3.6~\mathrm{cm}}$=0.0$\pm$0.5
(1$\sigma$ error; where $S_\nu$ $\propto$ $\nu^{\alpha}$), which is indicative of optically thin free-free emission (typically arising in shocks for the case of low-mass young stellar objects and VeLLOs, see \S~\ref{sdj041757protoBD}).
%The flux density at 6 cm was already given in  Barrado et al. (2009), nevertheless emission at 3.6 cm was not reported because of technical problems of the array during the observations at that frequency. In this work, we recalibrated and reanalyzed carefully those data. We identified high closure errors due to non-matched bandpass shapes between VLA and EVLA antennas, which were successfully fixed in the data reduction process and allowed us to detect marginally continuum emission at 3.6~cm (see \S 2 for more details).
To extract the morphological information of the centimetre emission, we imaged the B-configuration data only, and the resulting images are presented in Figure~\ref{fvla}. The figure shows first that the centimetre source is clearly associated with the NIR source `B'  detected by Barrado \et\ (2009), while source `A' shows no centimetre emission.
Furthermore, the figure reveals at both 6 and 3.6~cm a faint extension in the southwest-northeast direction. A Gaussian fit to the source yields, at both wavelengths, a deconvolved size of approximately $\sim2.5\times1.0$~arcsec$^2$, oriented at PA$\sim 40$~\degr\ (Table~\ref{tvla}).

%In order to improve the aspect of the maps, we have combined both data sets at different spatial resolution.  

%The flat spectral index value (compatible with optically thin free-free emission) and the elongated structure observed at higher angular resolution observations suggest that the radio continuum emission centreed at the position of the source J041757, could arise from an ionized jet.  More sensitive interferometric observations must be done in order to confirm this hypothesis.
 
 \begin{table}
\caption{VLA radio continuum emission detected at 3.6 and 6 cm towards J041757}
\begin{center}
{\small
\begin{tabular}{cllccc}
\noalign{\smallskip}
\hline\noalign{\smallskip}
$\lambda$&
\multicolumn{2}{c}{Position$^\mathrm{a}$}
%Position$^\mathrm{b}$&
%\colhead{Position\tablenotemark{a}}&
&$S_\nu^\mathrm{b}$&
Size, P.A.$^\mathrm{a}$&
\\
\cline{2-3}
(cm)&  
$\alpha (\rm J2000)$&
$\delta (\rm J2000)$&
(mJy)&
(arcsec$^2$, \degr)\\ 
\noalign{\smallskip}
\hline\noalign{\smallskip}
6     		&04:17:57.74  &$+$27:41:04.2       &0.20$\pm$0.05	&$2.4\times1.3,\, 45$\\
3.6     	&04:17:57.75  &$+$27:41:04.4       &0.19$\pm$0.11	&$2.9\times0.8,\, 40$\\
%4.86     &04$^{h}$17$^{m}$57$\fs$750  &$+$27$\degr$41$\arcmin$04$\farcs$40    &0.09    &0.20$\pm$0.03\\
%8.44     &04$^{h}$17$^{m}$57$\fs$740  &$+$27$\degr$41$\arcmin$04$\farcs$41    &0.10    &0.19$\pm$0.05\\
\hline
\end{tabular}
\begin{list}{}{}
\item[$^\mathrm{a}$] Results based on observations in B configuration of the VLA (see \S~\ref{sovla}), and derived from Gaussian fits. Uncertainties in sizes are around $0.5$~arcsec and in position angles (P. A.) are around 10\degr. Sizes and P. A. are deconvolved.
\item[$^\mathrm{b}$] Results based on observations in D configuration of the VLA. Uncertainties are 2$\sigma$.
\end{list}
}
\end{center}
\label{tvla}
\end{table}
%D CONFIG: banda C (04 17 57.99, 27 41 04.9) +/- 1.2"  1 sigma
%	   banda X (04 17 57.75  27 41 03.0) +/- 1.3"  1 sigma
%Combinacion sin pesado:
%	   banda C (04 17 57.740 27 41 04.41)   +/- 0.09 (1 sigma)
%	   banda X (04 17 57.75  27 41 04.4) +/- 0.1 (1 sigma)
% He usado ls posiciones de la combinacion sin pesado (salen muy parecidas en conf. D, B, y combinando con pesado), y los flujos de la configuracion D.	
%\tablenotetext{a}{Absolute position error.} 

\subsubsection{IRAM\,30m molecular emission \label{sremir}}

\begin{figure}
\begin{center}
\begin{tabular}[b]{c}
 \epsfig{file=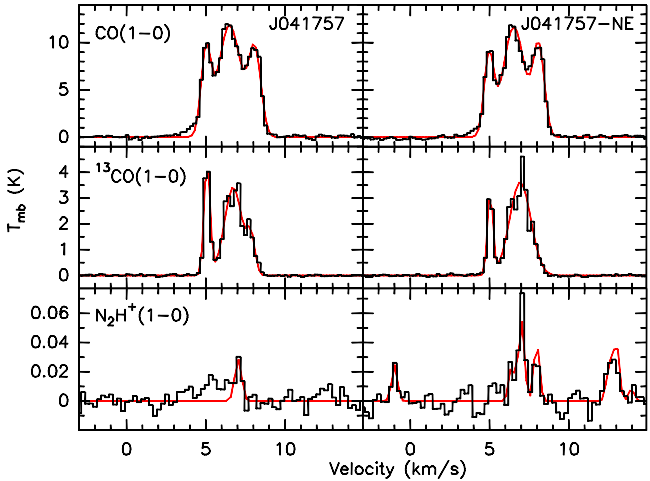, width=8.5cm, angle=0}\\
\end{tabular}
\caption{ 
Left column: spectra averaged in a region of $\pm20$~arcsec with respect to the position of J041757. Right column: idem for J041757-NE (offset of (37,30)~arcsec with respect to J041757). Note the excess of blueshifted emission at velocities in the range 2--4~\kms\ for CO\,(1--0).  
%Note also that the component around 5~\kms\ is stronger for J041757 than for J041757-NE, while the component at $\sim7$~\kms\ is stronger in J041757-NE.
For the case of \nth, J041757 was fitted with one Gaussian, while for J041757-NE we did fit the hyperfine structure of \nth.
}
\label{fspecs}
\end{center}
\end{figure}

\begin{figure*}
\begin{center}
\begin{tabular}[b]{c}
 \epsfig{file=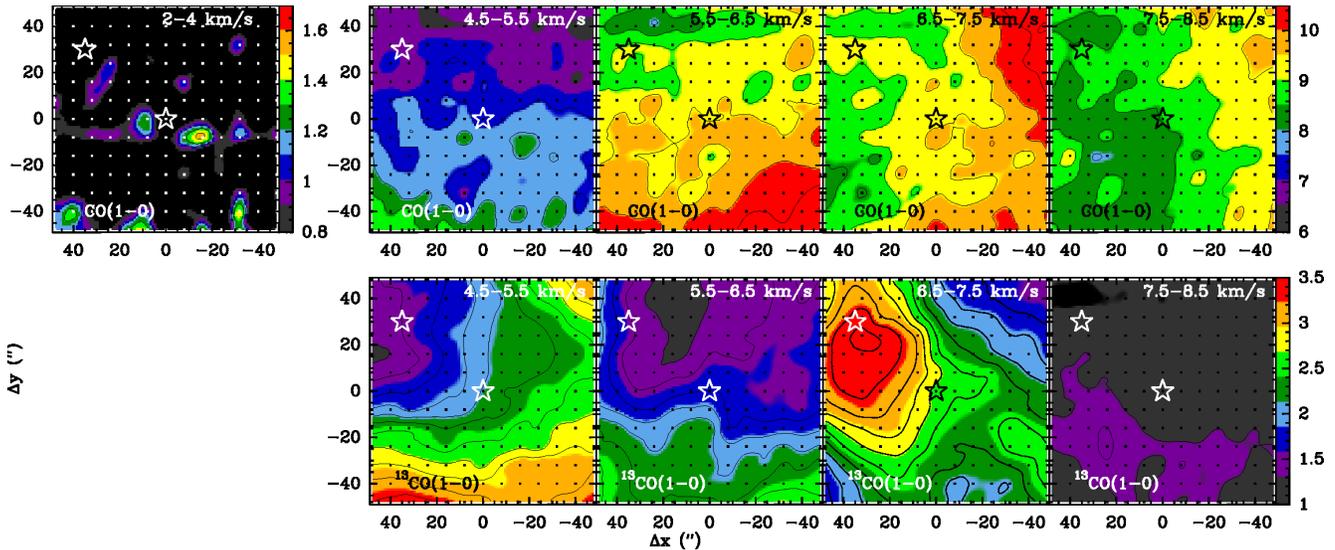,  width=17.5cm, angle=0} \\
\end{tabular}
\caption{ 
{\bf Top:} \co(1--0) channel maps in the J041757 region. Rms is 0.16~K (or 0.23~K for the panel showing the emission integrated from 2--4~\kms).
{\bf Bottom:} \tco(1--0) channel maps in the J041757 region.  Rms is 0.05~K.  For all panels, colorscale is in main beam brightness temperature (K), the beam is $\sim22$~arcsec, and stars mark the position of J041757 and J041757-NE.
}
\label{fchmaps}
\end{center}
\end{figure*}

The CO\,(1--0) spectra averaged over a region of 40~arcsec of diameter around J041757 (left) and J041757-NE (right) are shown in Fig.~\ref{fspecs}-top. The spectra reveal the three well-known velocity components associated with the Taurus molecular cloud, at around 5, 7 and 8~\kms\ (\eg\ Heyer \et\ 1987; Mizuno \et\ 1995).  
Among the three main velocity components seen in Taurus, the velocity maps of Fig.~6 of Mizuno \et\ (1995) show that at the position of J041757 there are two intersecting filamentary structures with different velocities: one in the 4.5--6.0~\kms\ range, and the other in the 6.0--7.5~\kms\ range. We made line area maps by integrating the CO emission in ranges of 1~\kms\ wide, from 4.5 to 8.5~\kms, and found that for this velocity range CO traces the large-scale structure of the B211 filament (located to the south-west of J041757, see Fig.~\ref{fextinction}), with strong intensity gradients and no clear features associated with J041757 (Fig.~\ref{fchmaps}-top). This was expected because CO is optically thick and spread out over all the Taurus region (\eg\ Goldsmith \et\ 2008; Narayanan \et\ 2008). 
However, we fitted with gaussians the three velocity components seen in the CO spectrum and found, 
for the emission line at 5~\kms, an excess in the blueshifted emission with respect to the gaussian fit\footnote{An excess with respect to a Gaussian fit was also found with the Kitt Peak Telescope in \co(1--0) between 2 and 4~\kms\ (O. Morata, priv. commun.), with the integrated emission also concentrated on J041757.}.
The line area map integrated in the range 2--4~\kms, where the blueshifted excess is seen, indicates that this emission is strongest around J041757, suggesting a possible relationship (Fig.~\ref{fchmaps}-top). 
%However,  such wings have been commonly found in CO in Taurus, and have been attributed to turbulence (\'Alvaro H., Mario: REFS????).

Concerning \tco(1--0), well suited to trace the gas column density in the region because is optically thinner than \co(1--0), the averaged spectra also reveal at least 3 velocity components for both J041757 and J041757-NE. In Fig.~\ref{fchmaps}-bottom, we show the line area maps of the \tco\ emission integrated between 4.5 and 8.5~\kms\ in steps of 1~\kms. Among the four mapped velocity intervals, the emission integrated between 6.5 and 7.5~\kms\ is more prominent at the position of J041757-NE, and has an extension to the south and another extension or stream to the southwest, passing through J041757. This could be a hint of the molecular gas associated with J041757-NE and J041757 coming from the same physical region.

Finally, the \nth(1--0) averaged spectra for J041757 and J041757-NE reveal emission associated with J041757-NE (see Fig~\ref{fspecs}-bottom). We fitted the hyperfine structure of the \nth(1--0) transition for J041757-NE using the package CLASS of the GILDAS software, and obtained the output parameter $A\tau= 0.059\pm0.006$~K\footnote{In the CLASS package of the GILDAS software, the first output parameter of a fit to the hyperfine structure of a given transition is $A\tau\,=f\,(J_\nu(T_\mathrm{ex})-J_\nu(T_\mathrm{bg}))\tau$,
where $f$ is the filling factor, assumed to be 1, $J_\nu(T)=h\nu/k/(e^{h\nu/kT}-1)$, with $k$ being the Boltzmann constant, T the temperature and $\nu$ the frequency; $T_\mathrm{ex}$ is the excitation temperature, $T_\mathrm{bg}$ the background temperature, and $\tau$ is the opacity of the main hyperfine component, $F'_1F'$--$F_1F=$23--12 for the case of \nth(1--0).}, 
$V_\mathrm{LSR}=7.02\pm0.02$~\kms, $\Delta v=0.42\pm0.17$~\kms, and an opacity of the $F'_1F'$--$F_1F=$23--12 transition of $\sim0.1$, which corresponds to an excitation temperature of 3.4~K and a \nth\ column density  of $2.5\times10^{11}$~cm$^{-2}$. This corresponds to a total H$_2$ column density of $\sim9\times10^{20}$~cm$^{-2}$ (assuming a \nth\ abundance of (2--5)$\times10^{-10}$, Pirogov et al. 2003, Tafalla et al. 2004), and a mass of $\sim 3$~\mj\ (for a size of $\sim20$~arcsec in diameter), similar to the lower limit of the mass derived from the submillimetre continuum emission (Table~\ref{tcso}). Towards J041757, the strongest hyperfine component of the \nth(1--0) transition is detected (at 5$\sigma$), and a gaussian fit yields  $7.04\pm0.08$~\kms, very similar to the velocity of the J041757-NE source. This suggests that both objects are associated with dense material (as the critical density of \nth(1--0) is $\sim10^5$~\cmt) possibly coming from the same region of Taurus, as indicated also by the morphology of the \tco\ emission integrated between 6.5 and 7.5~\kms\ (Fig.~\ref{fchmaps}-bottom).
%See the figure where a plot is shown for the \nth\ emission integrated from 6.3 to 7.3~\kms, revealing that the strongest \nth\ in J041757-NE extends to the southwest up to the J041757, where the emission is fainter but still stronger than in the surroundings (Fig.~\ref{fmolecmaps}).

\section{Spectral Energy Distributions (SEDs)}
\label{saseds}

\begin{figure}
\begin{center}
\begin{tabular}[b]{c}
 \epsfig{file=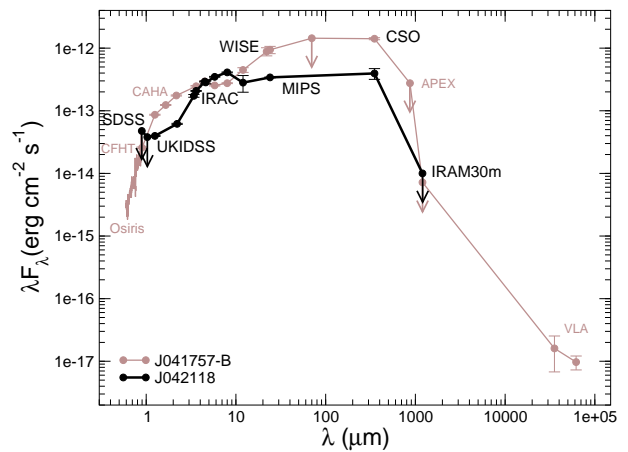, width=7.5cm, angle=270}\\
 \end{tabular}
\caption{Spectral Energy Distribution of J042118 (black line) and J041757, NIR component B (brown line), after Barrado \et\ (2009). 
%We fitted a modified blackbody with $\Td\sim10$~, $M_\mathrm{disc}\sim3$, 7 and 20~\mj, and dust emissivity index around 2.0, and used the opacity law of Ossenkopf \& Henning (1994).  
}
\label{fseds}
\end{center}
\end{figure}

Both J042118 and J041757 are Spitzer/IRAC sources emitting at 350~\mum. 
In order to further constrain its nature, we checked the NASA Extragalactic Database (NED), the National VLA Sky Survey (NVSS) at 21~cm, and JCMT/SCUBA catalog at 450 and 850~\mum\ (di Francesco \et\ 2008),  and found no data for SCUBA, no significant emission in NVSS above 1~m\jpb, and no registered object in NED. 
%Thus, there is no obvious reason to think that J042118 and J041757 must be discarded as Taurus members. 
%
To look for further archival data we used the new capabilities of the Virtual Observatory SED Analizer (VOSA, Bayo \et\ 2008; Bayo \et\ 2012 in prep.). The tool queries virtual observatory compliant catalogs in the ultraviolet (GALEX, IUE HPDP photometry), optical (Tycho-2, CMC-14, Hauck \et\ 1997, Mermilliod \& Mermilliod 1994, Sloan DSS release 7, and IPHAS), near, mid and far-infrared regimes (2MASS, IRAS, MSX6C, AKARI IRC and FIS, GLIMPSE, UKIDSS DR7, WISE preliminary release), and allows to fit several grids of models  with extinction as a free parameter. Appart from the data already presented in this work and in Barrado \et\ (2009), we find UKIDSS counterparts for both sources but with unreliable photometry in the H band, and WISE counterparts at less than 1~arcsec from the Omega2000 position of J041757 (NIR component B) and the Spitzer/IRAC position of J042118. Since Sloan DSS did not detect any source at the position of J041757 and J042118, we give only upper limits at the corresponding frequencies. In the Appendix we list the flux density measurements and upper limits (Table~\ref{tsedj042118} and \ref{tsedj041757b}), and in Fig.~\ref{fseds} we present the SEDs for both J042118 and J041757.

\section{Discussion}

\subsection{On the nature of J042118 \label{sdj042118nature}}

J042118 is undetected in Sloan DSS (data release 8) but detected in UKIDSS (data release 7, Fig.~\ref{fcso}), WISE, Spitzer/IRAC+MIPS and the CSO (see Table~\ref{tsedj042118}). 
In order to test if the J042118 SED (\S~\ref{saseds}) is easily reproducible by a very extincted photospheric model (as expected if it was a background object seen through the molecular cloud), we used VOSA tool (\S~\ref{saseds}) and found that even allowing extinction values up to $A_\mathrm{v}\sim15$~mag, it is not possible to get a reasonable fit (reduced $\chi^2$ values of the order of hundreds) neither with Kurucz (Castelli et al 1997) nor with Lyon models (Allard et al 1997; Baraffe et al 1997, 1998; Hauschildt et al 1999; Chabrier et al. 2000; Allard et al 2001; Allard et al 2003, 2007, 2009), suggesting that the object belongs to Taurus.

\begin{figure}
\begin{center}
\begin{tabular}[b]{c}
 \epsfig{file=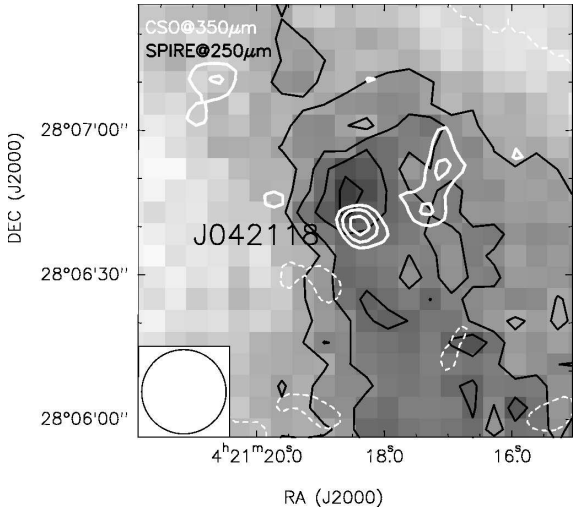, width=7.5cm, angle=0}\\
 \end{tabular}
\caption{
Submillimetre emission towards J042118. Greyscale and black contours correspond to the Herschel/SPIRE 250~\mum\ emission. 
Contours range from 0.06 to 0.12, increasing in steps of 0.02~\jpb\  (1$\sigma$ is approximately 0.009~\jpb). Herschel beam at 250~\mum, shown in the bottom-left corner, is $\sim17$~arcsec. White contours are the CSO 350~\mum\ emission (same contours as in Fig.~\ref{fcso}), and CSO beam is $\sim10$~arcsec.
}
\label{fherschel_j042118}
\end{center}
\end{figure}

With the aim of shedding light on the nature of J042118, we compared the CSO emission to the Herschel Archive images at 250~ and 160~\mum\footnote{
Herschel Archive data from the `Herschel Gould Belt survey'  Key Program (see Andr\'e \et\ 2010) were downloaded (observing identification numbers 1342202090 and 1342202254). The Herschel/PACS(SPIRE) data at 160(250)~\mum\ were processed with HIPE(madMap) version 8 and Scanamorphos version 14 (Roussel 2012).
The Herschel data towards B211/L1495 will be published in a forthcoming paper by Palmeirim \et\ (in prep.) as part of the `Herschel Gould Belt survey' project.
}, 
which are the wavelengths for which the Herschel beam (17~arcsec and 11~arcsec) is more similar to the CSO beam (10~arcsec). 
%
%The Herschel data towards B213/L1495 belong to the `Herschel Gould Belt survey' Key project (Andr\'e \et\ 2010) and will be published in a forthcoming paper by  Palmeirim \et\ (in prep.).
%
While at 160~\mum\ there is no significant emission towards J042118, the Herschel/SPIRE 250~\mum\ emission shows an extended large-scale structure elongated in the southwest-northeast direction (see Fig.~\ref{fherschel_j042118}), with a subcondensation matching well the position of J042118, within the telescope positional uncertainties.
%Herschel peak is $\sim4''$ to the east with respect to the CSO emission. 
%The flux density measured by Herschel at 250~\mum\ is consistent with the flux density measured by CSO, assuming a power-law index of 3. .
The location of J042118 at the tip of the elongated structure, which seems to be part of the extended structures of the Taurus complex, favors its membership to Taurus.
In this case, the submillimetre source detected with the CSO would be tracing a disc and/or small ($<5$~arcsec or 700~AU of radius) envelope with a mass of $\sim1$~\mj\ (Table~\ref{tcso}).
%A first analysis of the Spectral Energy Distribution (SED) for J042118 (Fig.~\ref{fsedj042118}, Table~\ref{tsedj042118}) suggests a mass of the envelope of 3~$M_\mathrm{Jup}$, a dust temperature of 12~K, assuming a dust emissivity index of 2 (and an opacity at 300 micron of 0.123 XXX from Ossenkopf \& Henning 1994).
The luminosity (upper limit) for J042118 estimated from the SED (Table~\ref{tsedj042118}) would be $\la0.0023$~\lo, and the SED shape is reminiscent of Class I(/II) young stellar objects, with a flat slope between the NIR and the submillimetre range.
%(different from J041757, which peaks at around 100~\mum\ and shows extended emission, as young stellar objects in the Class 0 stage).
This can be quantified by estimating the bolometric temperature, of $\sim 140$~K\footnote{This number must be regarded with caution because the SED for this source is better sampled in the IR than in the submillimetre and radio range, and because we treated all flux densities which are upper limits as detections (c.f. Kauffmann \et\ 2008).}, typical of Class I young stellar objects (\eg\ Chen \et\ 1995). 
% 2012Mar7:
% Lbol=  0.00229721278Lsun
% Tbol=  143.179661K
The fact that the submillimetre emission, as recovered by the CSO, is unresolved could be indicative of most of the envelope being accreted/dispersed, with the submillimetre emission arising only from a compact disc. 
%However, it should be confirmed that the object belongs to Taurus to definitely classify it as a BD in the Class I/II stage.

%Herschel towards J042118:
% PACS160mic: no detection
% SPIRE250mic: nice "stream" again with a peak at J042118 :)  -> good to show in the paper
% SPIRE350mic: again the stream, but the peak is not so clearly associated with J042118
% SPIRE500mic: idem as 350mic

\subsection{On the nature of J041757 \label{sdj041nature}}

The nature of J041757 is discussed in Barrado \et\ (2009), who present series of color-color and color-magnitude diagrams from Spitzer and NIR data to conclude that the most likely scenario for J041757 (for either component A and/or B) is that the object is galactic. Subsequent infrared spectroscopic observations by Luhman \& Mamajek (2010) indicate that component A is most likely a M2 background star, while no conclusion could be drawn for component B because no spectrum was shown for this component in that work. This, together with the fact that source B has a raising slope in the NIR SED matching well the FIR and submillimetre emission from Spitzer/MIPS and CSO (Barrado \et\ 2009), indicates that the most likely counterpart of the submillimetre source in J041757 is the NIR source B, which in turn is associated with centimetre emission (\S~\ref{srvla}). Hereafter we will refer to source B (identified in the NIR and in the centimetre range) as J041757-B.

\begin{figure}
\begin{center}
\begin{tabular}[b]{c}
    \epsfig{file=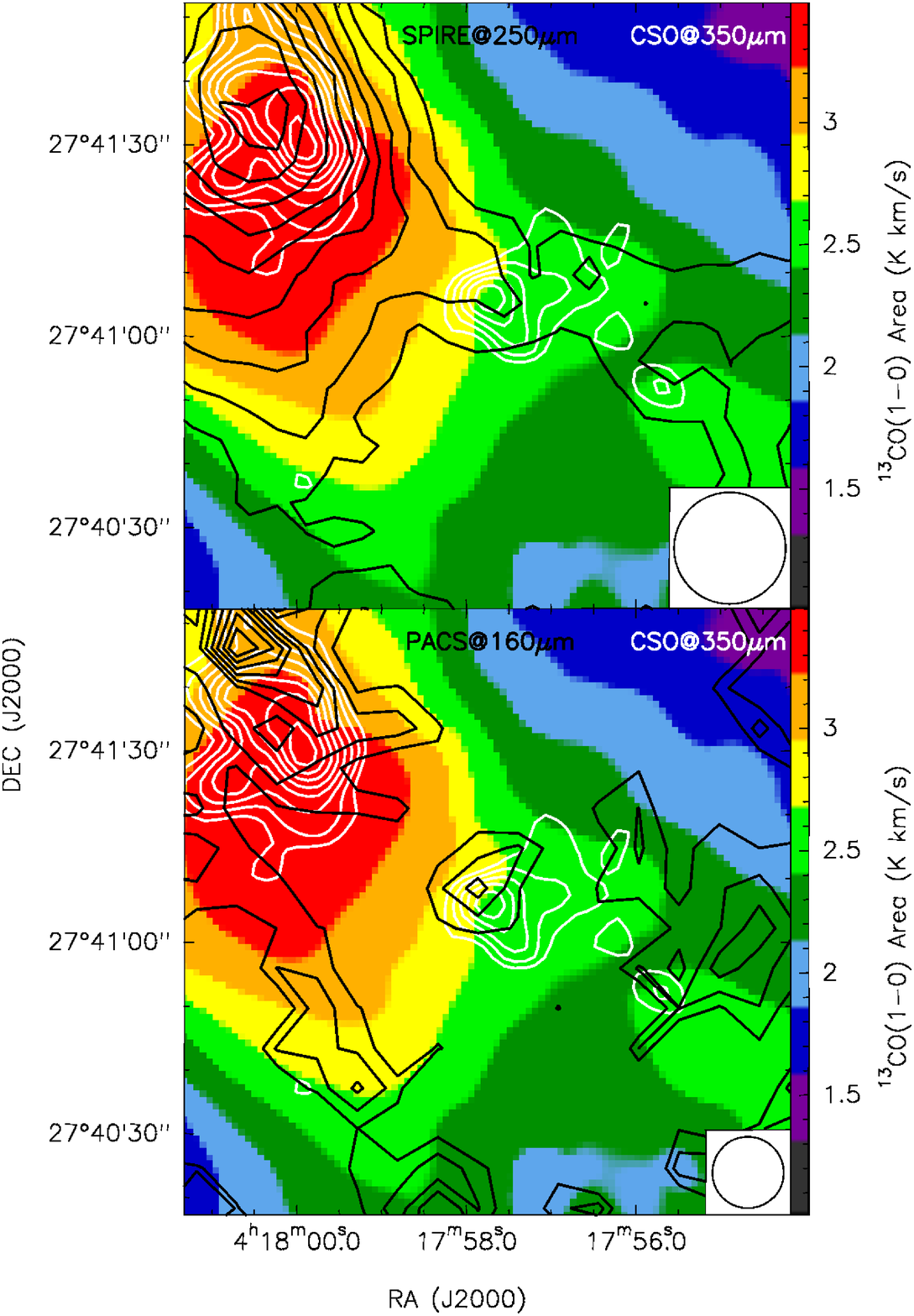, width=8.5cm,angle=0}\\
\end{tabular}
\caption{
J041757: in both panels, colorscale is the \tco(1--0) area integrated in the range 6.5--7.5~\kms\ (as in Fig.~\ref{fchmaps}) and white contours are the CSO 350~\mum\ emission (same contours as in Fig.~\ref{fcso}).
{\bf Top:} Black contours correspond to the Herschel/SPIRE 250~\mum\ emission. 
Contours range from 0.30 to 0.54, increasing in steps of 0.03~\jpb\ (1$\sigma$ is approximately 0.014~\jpb). Herschel beam at 250~\mum, shown in the bottom-right corner, is $\sim17$~arcsec (\tco\ beam is $\sim22$~arcsec).
{\bf Bottom}: Black contours correspond to the Herschel/PACS 160~\mum\ emission. 
Contours range from 0.14 to 0.24, in steps of 0.014~\jpb\ (1$\sigma$ is approximately 0.018~\jpb). Herschel beam at 160~\mum, shown in the bottom-right corner, is $\sim11$~arcsec, similar to the CSO beam.
%0.1367 to 0.2374, in steps of 0.014~\jpb\ (1$\sigma$ is approximately 0.018~\jpb).
%Contours range from 0.038 to 0.066, in steps of 0.004~Jy/px.
}   
\label{fherschel}
\end{center}
\end{figure}

Similarly to the case of J042118, we compared the CSO emission of J041757 to the Herschel Archive images at 250~ and 160~\mum. See Fig.~\ref{fherschel} to visualize such a comparison. 
%For the case of J042118, the Herschel submm emission shows an extended large-scale structure which is strongest and widest to the southwest of J042118, and extends for about $2'$ to the northeast, progressively decreasing its intensity ang getting narrower. Such a large-scale submm structure shows a subcondensation matching well the CSO compact source in J042118, except that the Herschel peak is $\sim4''$ to the east with respect to the CSO emission. 
The Herschel/SPIRE emission at 250~\mum\ (black contours in Fig.~\ref{fherschel}-top) is strongest at the position of J041757-NE, matching well the 350~\mum\ emission observed with the CSO (white contours). From J041757-NE, there are two streams of emission, extending for about $\sim1$~arcmin, one towards the south and the other towards the southwest and passing through J041757-B. The 250~\mum\ emission towards the southwestern side of the map is part of the B211 filament (Fig.~\ref{fextinction}).
We note the remarkable similarity between the emission at 250~\mum\ and the \tco(1--0) emission integrated between 6.5 and 7.5~\kms\ (colorscale in Fig.~\ref{fherschel}-top). Compared to Herschel data, the CSO is recovering the most compact (as we used the `compact' option in the reduction process, see \S~\ref{socso}) and fainter emission (1$\sigma$ noise of Herschel/SPIRE at 250~\mum: $\sim14$~m\jpb; 1$\sigma$ noise of CSO at 350~\mum: $\sim6$~m\jpb). Regarding the emission seen with Herschel/PACS at 160~\mum, this is again strongest at J041757-NE, and the northeast-southwestern stream splits up into one slightly resolved source coinciding,  within the position uncertainties of the telescopes, with the 350~\mum\ source at the position of J041757-B.
%(the small offset of about $\sim5$~arcsec between the 160~\mum\ source and the 350~\mum\ CSO source is within the positional uncertainties of both telescopes). 
This suggests that J041757-B is associated with the hottest dust of the stream linking the J041757-NE source and the B211 filament.
%J041757 detected at 160~\mum\ (PACS) at a level of 48~m\jpb, while the nominal point-source sensitivity of PACS at 160~\mum\ is $\sim 10$~mJy (Poglitsch \et\ 2010), which would imply that J041757 would be detected at around 5$\sigma$ (Fig.~\ref{fpacscso}).
Flux densities measured by Herschel at 160 and 250~\mum\ are consistent with the flux densities given here as measured with the CSO, but we refrain from presenting a detailed study of the Herschel data in this work as this will be the subject of a forthcoming paper.

\subsubsection{J041757-B: an extragalactic object? \label{sdj041757extragal}}

%There is no identified known extragalactic object at this position as catalogued in the NASA/IPAC Extragalactic Database.
%In addition, the mag-color and color-color diagrams for this object show that it is not fully consistent with the nature of spiral galaxies, AGN's, and QSOs, indicating that if the object is extragalactic indeed, its nature is really intriguing (Barrado \et\ 2009).
%Barrado et al. (2009), after carefully checking mag-color and color-color diagrams using Ic, JHK and Spitzer-IRAC and MIPS filters (Figs. 4 and 5 of that paper), conclude that J041757-B photometry in these frequency ranges does not match with any particular galaxy type among AGNs, QSOs, and star-forming galaxies (SFGs)????. For example, if the J041757-B has typical colors of AGN's in the XXXXX color-color diagram, then in the XXXX diagram it is not compatible with being an AGN. The same applies for QSOs and SFGs (RECHECK THIS AGAIN!!!!).

\begin{figure}
\begin{center}
\begin{tabular}[b]{c}
 \epsfig{file=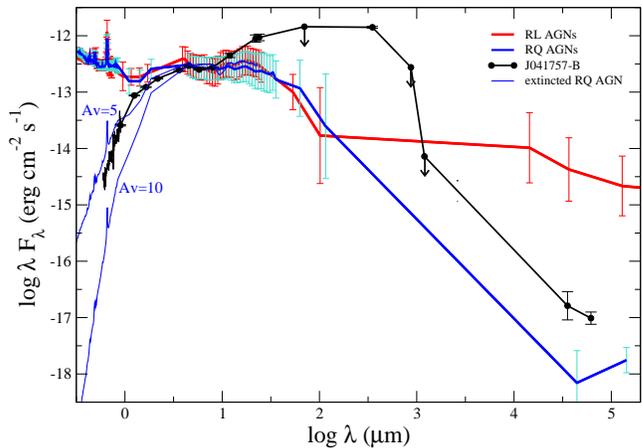, width=7.5cm, angle=270}\\
\end{tabular}
\caption{Spectral Energy Distribution of J041757-B (black) compared to average SEDs for radio-loud (thick red) and radio-quiet (thick blue) AGNs from Shang \et\ (2011).  Thin blue lines are the extincted radio-quiet SEDs with $A_\mathrm{v}$ of 5 and 10~mag. The spectrum of J041757-B in the visible will be presented in a forthcoming paper (N. Hu\'elamo et al., in prep.).}
\label{fAGNs}
\end{center}
\end{figure}

We consider first the possibility that J041757-B is an extragalactic object, which would be aligned by chance with the submillimetre stream linking the J041757-NE source and the B211 filament.  In Fig.~\ref{fAGNs} we show a comparison of the SED for J041757-B to the recent compilation of average SEDs of radio-loud and radio-quiet AGNs of Shang \et\ (2011, which are in good agreement with the classical SEDs of Elvis \et\ 1994). The SEDs are normalized at 12~\mum\ (a good intrinsic indicator for AGNs, \eg\ Gandhi \et\ 2009) to the flux of J041757-B, and the error bars are the standard deviation of the fluxes for Shang et al. sample, being representative of the scatter in the AGN fluxes at each wavelength. From the figure, it is seen that J041757-B SED is not compatible with the radio-loud AGN SEDs, mainly because of its important drop from MIR to radio wavelengths\footnote{We note however that if we estimate the radio-loudness from the classical definition, $S_\mathrm{6cm}/S_\mathrm{4400\AA}$  (following Kellermann et al. 1989), we obtain $\sim700$, which would classify J041757-B as a radio-loud. As explained in the text, this would be still compatible with a radio-quiet AGN if the extinction for J041757-B is important.}.
%Most AGNs are radio-quiet (Barvainis \et\ 2005). 
In the visible range, J041757-B SED is much more obscured than AGN SEDs. In order to simulate an AGN extincted SED we reddened the AGN average SED for radio-quiet using the extinction law by Fitzpatrick (1999; improved by Indebetouw et al (2005) in the infrared, and $R_\mathrm{v}\sim3.1$), and with $A_\mathrm{v}$ ranging from 1 to 10 (see figure, thin blue lines). 
%Amelia report about reddening AGN SEDs:
%I took the SED that Daniel gave us and changed it to the resolution of the extinction law that I had that is only important up to ~30 micron; for redder wavelengths it is completely negligible (the extinction law reference is: extinction law by Fitzpatrick (1999) improved by Indebetouw et al (2005) in the infrared). I then reddened the sED for different values from 1mag to 10 mag as you can see in the fig. In the fig I am plotting lambda vs lambdaFlambda, but in the files that I am sending you, the x and y columns are really lambda Flambda (as indicated). At some point I changed from Jy to erg/cm2/s/Hz without multiplying by the units factor and that is why everything is in arbitrary units. The IDL script that I attach created the individual reddened SED files, if you want to play with other values of extincton just edit the Av vector (and the avstring one for the figure). 
The result is that the J041757 SED would be compatible with an AGN SED extincted by $A_\mathrm{v}\sim5$ in the visible-IR range. 
However, the J041757-B SED is not fully consistent with radio-quiet AGNs either, because of its important contribution of submillimetre (and radio) emission compared to MIR.  This could be indicating that there is an excess of submillimetre emission in the direction of the galaxy, which could be explained by the Taurus stream joining the J041757-NE source and the B211 filament. Nevertheless, there would be an excess of submillimetre emission from the galaxy candidate itself, as revealed by the image of Herschel/PACS at 160~\mum, which shows that in the stream there is only emission at 160~\mum\ at the position of J041757-B (similar case for the CSO 350~\mum\ emission, Fig.~\ref{fherschel}). In addition, flat spectral indices in the centimetre range in AGNs are not usual, as centimetre emission from AGNs is typically produced by synchrotron emission from relativistic electrons accelerated in jets (\eg\ Giroletti \& Panessa 2009; Ibar \et\ 2009; Shang \et\ 2011; Randall \et\ 2012). Furthermore, a preliminar spectrum of J041757-B in the visible range taken with Osiris on the GTC (Hu\'elamo et al., in preparation) shows no broad emission lines, while the extincted SED in Fig.~\ref{fAGNs} indicates that even with $A_\mathrm{v}\ga5$ still strong and broad lines are expected for an AGN. If J041757-B was a galaxy it should be detected in forbidden emission lines in the MIR range ([OIV], [NeV]: \eg\ Goulding et al. 2009) and in hardest X-rays ($>10$~keV)\footnote{In X-rays, we could only set an upper limit for J041757-B, which is not detected in the ROSAT All-Sky Survey (RASS, Voges et al. 1999, covering the range 0.1--2.4~keV with a typical limiting sensitivity of $10^{-13}$~erg\,cm$^{-2}$\,s$^{-1}$, Anderson \et\ 2003).}, which could be the final observational tests to definitively discard the nature of J041757-B as a background galaxy.

\subsubsection{J041757-B: a background object behind Taurus? \label{sdj041757galback}}

Similarly to J041757-A (Barrado \et\ 2009, Luhman \& Mamajek 2010), J041757-B could be a background object behind the Taurus complex. 
However, if
%after Barrado \et\ (2009) SEDs of both J041757-A and J041757-B (with J041757-A strongly decreasing from 1.6 to 4.6~\mum), and the spectroscopic observations of Luhman \& Mamajek (2010), it seems that 
J041757-A is a background star, the IR source mainly contributing to the FIR and submillimetre emission must be J041757-B.
%IRAC emission at 5.8 and 8~ \mum, and to the MIPS emission at 24~\mum, must be J041757-B, which in turn must be the object associated with the CSO source at 350~\mum, (Herschel-PACS source at 160~\mum) and also to the cm source at 6 and 3.6 cm. 
Therefore, whatever the true nature of J041757-B is, the object is embedded in considerable amounts of dust, and hence should be associated with molecular gas as well.  This molecular gas, if J041757-B does not belong to Taurus, should appear as a velocity component in the line of sight different from Taurus velocities, while in a 9000 km/s band we did not detect any other line apart from those of Taurus. If the object is associated with dust and gas, it should be a young stellar object (most likely in the Class II/III phase) or an evolved star.  However, the centimetre emission for Class II/III objects is typically due to non-thermal gyrosynchrotron emission, which is not resolved (\eg\ Andr\'e \et\ 1992, Forbrich \et\ 2007) and is polarized, while J041757-B has a flat (thermal) spectral index, is slightly elongated in the southwest-northeast direction, and shows no polarized emission (we checked this in the dataset of 2008 observations and had no detection of polarized emission). In addition, the lack of X-ray emission (no ROSAT detection) is also uncommon for Class II/III objects (\eg\ G\"udel et al. 2007).
%BUT RASS SENSITIVITY IS WORSE THAN Guedel et al. 2007:
%Guedel et al. 2007: detection threshold of ~10^28 erg s?1
%ROSAT All-Sky Survey: typical limiting sensitivity of $10^{-13}$~erg\,cm$^{-2}$\,s$^{-1}$:
%Corresponds to Lx ~10^31 at 1 kpc, or 10^32 at 3 kpc
%While typical Lx for Class II are: 10^28 to 10^32
% Guedel et al. 2007: In this study it is shown that X-ray detection is usual for Class II objects (only 15\% undetected), but not for protostars (60\% undetected) (cited in Scholz \et\ 2010,  MNRAS, 409, 1557). 

The possibility that J041757 is an evolved stellar object, either a carbon star or a giant, is also considered. 
Luminous carbon stars are evolved stars with more carbon than oxygen in their atmosphere, characterized by red near-infrared colors and spectra with no \water\ (Liebert et al. 2000). The JHK colors for J041757-B match well within the JHK color-color diagrams for these stars (\eg\ Quanz \et\ 2010).  
However, we used the $K_\mathrm{s}$ magnitude of J041757-B (Barrado \et\ 2009) and Table~8 from Quanz \et\ (2010) to estimate the distance and find $>160$~kpc, larger than the currently known extent of our Galaxy. Since the emission in J041757 is resolved in the submillimetre range, such a large distance would imply an unrealistic large size for the envelope of a carbon star of $\sim9$~pc.
In addition, the carbon star SEDs have the peak at around 12~\mum, and longwards the flux decreases strongly (\eg\ Liebert \et\ 2000), which is clearly not the case of J041757-B. Thus, the properties of J041757-B are not fully consistent with being either a Class II/III young stellar object, or an evolved star.
In fact, by making use of the data compiled to build the SED (\S~\ref{saseds}), and using the VOSA tool 
as outlined in \S~\ref{sdj042118nature}, we found that the J041757-B SED cannot be reproduced with a very extincted photospheric model, providing further support against the galactic background object scenario for J041757-B.

\subsubsection{J041757-B: a proto-brown dwarf candidate in Taurus? \label{sdj041757protoBD}}

%One of the issues raised by Luhman \& Mamajek (2010) against the background nature of J041757-B is that the proper motions ???????? Further analysis confirmed the proper motions given in Barrado \et\ (2009), and J041757-B falls outside the main "cloud" traced by Taurus members but still is perfectly compatible with the proper motions of some Taurus members.
%(Table 3 and Sect. 5.2 of BERTOUT ET AL. 2006 list the Taurus members with no proper motion: not clear if these Taurus members share a common property...; Quanz et al. 2010, Fig.9: there are Taurus members with no proper motion, similar to J041757-B). 

Finally, J041757-B could be an embedded object belonging to Taurus. This is strongly suggested by the discovery of a narrow ($\sim16$~arcsec width) stream joining the J041757-NE source and the B211 filament and passing exactly through the position of J041757-B. Since both J041757-NE and the stream are detected at around 7~\kms\ (in \nth\ and/or \tco), both structures belong to Taurus.
%, while the expected number of galaxies (of $\sim0.2$~mJy at 6~cm) within a field of view of $16''$ is only $10^{-3}$ (following Anglada et al. 1998), implying a small probability of such a chance alignment.

If J041757-B belongs to Taurus, its low luminosity, of $\la0.005$~\lo\ (using the values in Table~\ref{tsedj041757b}), would imply a proto-BD nature (Barrado \et\ 2009). 
% 2012May20:
% Lbol=  0.00456646214Lsun
% Tbol=  122.308526K
This would naturally explain the centimetre emission with a flat spectral index, since flat/positive spectral indices, indicative of thermal free-free emission, are typically observed in young stellar objects due to shocks in the jet (\eg\ Anglada \et\ 1998; Beltr\'an \et\ 2001; Avila, Rodr{\'{\i}}guez, \& Curiel 2001; Pech \et\ 2010; Scaife \et\ 2011; Ward-Thompson \et\ 2011), and have also been found for Very Low Luminosity Objects (\eg\ Shirley \et\ 2007, Scaife \et\ 2011). In these cases, the gas is ionized in shocks produced by the jet driven by the nascent star, and the centimetre emission is usually elongated along the outflow axis. This would also easily explain the elongation seen at both 6 and 3.6~cm for J041757-B. And interestingly, the \co(1--0) excess detected between 2 and 4~\kms\ is reminiscent of wings observed in outflows driven by low-mass young stellar objects.
%Note that the VeLLO L1014-IRS also shows cm (thermal+non-thermal) emission (Shirley et al. 2007). Toward pre-protostellar cores only marginal/faint cm sources have been detected (Stamatellos et al. 2007).
%Stamatellos et al. 2007, A&A, 462, 677: folder: PPCs
In addition, if J041757-B is a proto-BD, its bolometric luminosity and centimetre luminosity match well the relation found for young stellar objects driving radiojets, proposed first by Anglada (1995). Following the recent correlations of Shirley \et\ (2007) and Scaife \et\ (2011) for the low-mass case, the expected centimetre luminosity, given the bolometric luminosity of J041757-B, is 0.0002--0.007~mJy\,kpc$^2$, consistent with the measured centimetre luminosity of $0.004\pm0.002$~mJy\,kpc$^2$. 
Last, if J041757-B belongs to Taurus, the properties of the extended envelope where the NIR source is embedded, of about $\sim1000$~AU of radius and about $\sim5$~\mj, coincide with the extrapolation of the trend presented  by Kauffmann \et\ (2008) in a radius versus mass diagram for small c2d Spitzer cores (see Fig.~4a of Kauffmann \et\ 2008), and more recently by Sadavoy \et\ (2010) and Schmalzl \et\ (2010) for the Taurus region.

Thus, if J041757-B belongs to Taurus, its properties can \emph{naturally} be explained simply as a scaled-down version of low-mass young stellar objects.

\subsection{On the nature of J041757-NE: a pre-substellar core? \label{sdnature}}

J041757-NE is the strongest submillimetre source detected in the field of J041757 with the CSO, and has been confirmed by APEX, IRAM\,30m, and Herschel images. This source is clearly associated with Taurus, as shown by the \nth\ and \tco\ emission detected at 7~\kms\ (Figs.~\ref{fspecs} and \ref{fchmaps}). Since we estimated a total mass (compact+extended emission) for J041757-NE of $\sim100$~\mj\ (\S~\ref{srj041}), and no point IR source (up to 24~\mum) has been found associated with the submillimetre peak,  J041757-NE could be harboring a true pre-substellar core, i.e., a starless core which will form a BD in the future. A first requisite for this to happen is that the submillimetre source should be gravitationally bound. Given the linewidth measured from the \nth(1--0) hyperfine fit (\S~\ref{sremir}), and an average radius of 10~arcsec (Table~\ref{tcso}), we estimate a gravitational mass of  140~\mj\ (following equation 4 of Pound \& Blitz 1993), similar to the total mass measured  for J041757-NE, suggesting that J041757-NE might be indeed a pre-substellar core. Interestingly, its total mass and radius follow the correlation found between mass and radius for protostellar/pre-stellar cores (\eg\ Kauffmann \et\ 2008, Sadavoy \et\ 2010; Schmalzl \et\ 2010). Also, the fact that this pre-substellar core candidate is found in B211, the portion of L1495 with smaller number of young stellar objects (\eg\ Schmalzl \et\ 2010) supports the idea suggested by Schmalzl \et\ (2010) that this region is at the very beginning of the star (or BD) formation process.

The number of pre-substellar core candidates in the literature is very scarce.
A pioneer study searching for pre-substellar cores and proto-BDs was carried out by Pound \& Blitz (1993, 1995), who 
%observed in C$^{34}$S and DCO$^+$ different positions in the Taurus and Ophiucus clouds, and 
found about 4 cores with substellar masses in the Taurus and Ophiucus clouds. However, the cores were at the detection limit and all but one were not gravitationally bound. In addition, the authors indicate that the candidate objects should be re-observed deeply to rule out the possibility that the cores extend considerably, making them massive enough to overcome the substellar domain.
More recent studies in Gould Belt clouds report samples of starless and pre-stellar cores, with some of them being substellar, but lack the kinematical information required to assess if they are gravitationally bound (\eg\ Ward-Thompson \et\ 2007; Davis \et\ 2010; Sadavoy \et\ 2010). 
Thus, J041757-NE  is an excellent pre-substellar core candidate, and high angular resolution millimetre observations would be extremely useful to study possible infall motions within the core, which would definitely indicate that the object is on the way to form a BD.\\

\section{Conclusions}

We present the results of a multiwavelength search for pre-substellar cores and proto-BD candidates in the B213-L1495 clouds of Taurus. The studied sample of 12 candidates, selected from Spitzer/IRAC photometry of the region (Barrado \et\ 2009), was observed at 1.2~mm with the IRAM\,30m/MAMBO-II bolometer. The two most promising objects, J042118 and J041757, were subsequently observed at 350~\mum\ with the CSO. Additional observations were performed toward J041757, to further constrain its nature. Our main conclusions can be summarized as follows:

\begin{itemize}

\item[(i)] None of the 12 Spitzer/IRAC selected sources observed with IRAM\,30m/MAMBO-II were detected at a level $>4\sigma$, and two sources, J042118 and J041757 had fluxes at the $2\sigma$ level. We set upper limits to the envelope masses of  $\la3$~\mj.

\item[(ii)] CSO 350~\mum\ observations towards J042118 reveal one compact submillimetre source associated with the selected Spitzer/IRAC source, with a size of $<10$~arcsec $\sim1400$~AU (at the Taurus distance), and a mass of $\sim1$~\mj. A preliminar analysis of the Spectral Energy Distribution indicates a luminosity of about 0.002~\lo, and a bolometric temperature around 140~K, typical of Class I/II young stellar objects.

\item[(iii)] CSO 350~\mum\ observations towards J041757 reveal one extended source associated with the selected Spitzer/IRAC source, of about $\sim16$~arcsec or $\sim2000$~AU of size, and a mass of $\sim5$~\mj. Additionally, we detect one sligthly resolved source at 3.6 and 6~cm exactly at the position of the infrared counterpart `B' of J041757, called J041757-B, with a flat spectral index. J041757-B is found at the centre of a stream of dust and gas, extending from the B211 main filament to the northeast, mainly detected with \tco(1--0) at a velocity of $\sim7$~\kms. \nth(1--0) emission is detected at the same velocity at 5$\sigma$ towards J041757-B. All this seems to indicate that J041757-B is probably a proto-BD in the Class 0/I stage associated with a radiojet and an extended envelope, whose properties seem to be a scaled-down version of low-mass protostars.
%whose properties match well the low-end of typical relations found for low-mass young stellar objects.

\item[(iv)] We discovered a partially extended submillimetre source about 40~arcsec to the northeast of J041757-B, of 20~arcsec $\sim3000$~AU of size and total mass of $\sim100$~\mj, called J041757-NE, and with no centimetre emission neither optical/infrared (up to 24~\mum) emission associated. This indicates that J041757-NE seems to be starless. J041757-NE is well detected in \tco\ and \nth, and is found also at $\sim7$~\kms. With the measured linewidth of \nth, and its measured size and mass, 
%The \tco(1--0) emission integrated in the range 6.5--7.5~\kms\ reveals a prominent clump of gas associated with J041757-NE, with a stream extending towards the southwest down to B213 and passing through J041757-B. J041757-NE shows also intense emission in \nth(1--0), for which we derived a velocity around 7~\kms, and a linewidth $\sim0.4$~\kms. Considering that possibly large-scale emission has been filtered out with the reduction of the 350~\mum\ data, we estimate that the true mass of J041757-NE could reach up to $\sim60$~\mj. With this mass estimate, and using the measured \nth\ linewidth and the submillimetre size, 
we conclude that J041757-NE is likely gravitationally bound and thus is a good pre-substellar core candidate.

\end{itemize}

Overall, 
%our multiwavelength study suggests a picture where a stream of the B213 filament, extending towards the northeast, is undergoing fragmentation, starting forming a proto-BD at the centre of the stream, and with a more massive pre-substellar core at the end of the stream. 
this work shows the power of submillimetre instruments to trace extremely low masses and thus to reveal substellar objects.
We found evidence of two proto-BD candidates and one pre-substellar core candidate which have scaled-down properties of low-mass young stellar objects and which are embedded in large-scale structures, all this favoring the {\it in-situ} scenario for BD formation in this portion of the Taurus complex.

%{\bf Therefore, in the J041757 area there could be two brown dwarf formation events, one which took place in the past and corresponding to J041757-B, and another event likely taking place in the future, corresponding to J041757-NE, with both objects forming part of the same elongated large-scale structure. 
%These results, together with the finding of J042118 embedded in a large-scale structure like J041757-B and J041757-NE, are difficult to reconcile with the ejection scenario for brown dwarf formation, and the discovered objects are rather ideal sites for further tests of brown dwarf formation as a scaled-down version of low-mass star formation.}

\section*{Acknowledgments}

We thank the referee, Thomas Henning, for key comments and suggestions.
A.P. is grateful to Attila Kovacs, Stephane Leon, Robert Zylka, and Jean-Fran\c cois Lestrade for
useful suggestions and help regarding the reduction of the CSO and
IRAM\,30\,m data, and to \'Alvaro Hacar, Mario Tafalla, Benjam\'{\i}n Montesinos, Rosario L\'opez, Mar\'{\i}a Rosa Zapatero-Osorio, 
Almudena Alonso, Giovanni Miniutti, Tommy Wiklind, and Violeta Gonz\'alez for insightful discussions. A.P. and N.H. are grateful to Catarina Alves de Oliveira, Florian Rodler, Markus Schmalzl, and Darren Dowel for kindly providing data for comparison, and D.B. is grateful to Manuel G\"udel for checking that the targets of this work have not been observed with Chandra and XMM.
We thank Calar Alto Observatory and IRAM\,30\,m Telescope for allocation of director's discretionary time to this programme.  
A.P. and I.dG-M are supported by the Spanish MICINN grant AYA2008-06189-C03 (co-funded
with FEDER funds) and A.P. is also supported by a JAE-Doc CSIC fellowship co-funded with the European Social Fund
under the program "Junta para la Ampliaci\'on de Estudios".
%IdG-M acknowledges partial support from Ministerio de Ciencia e Innovaci\'on (Spain), grant AYA2008-06189-C03-01.
%
This research has been partially funded by Spanish MICINN under the Consolider-CSD2006-00070,
AYA2010-21161-C02-02, and PRICIT-S2009/ESP-1496 grants,
% Amelia's:
and by the Marie Curie Actions of the European Commission (FP7-COFUND).
This publication makes use of VOSA, developed under the Spanish Virtual
Observatory project supported from the Spanish MICINN through grant
AyA2008-02156; of the SIMBAD database, operated at CDS, Strasbourg, France;
%This research has made use of the Spanish Virtual Observatory supported from the Spanish MEC through grant AyA2008-02156
%
of data products from the Wide-field Infrared Survey Explorer, which is a joint project of the University of California, Los Angeles, and the Jet Propulsion Laboratory/California Institute of Technology, funded by the National Aeronautics and Space Administration.
Some of the data reported in this paper were obtained as part of the United Kingdom Infrared Telescope (UKIRT) Service Programme. The UKIRT  is operated by the Joint Astronomy Centre on behalf of the Science and Technology Facilities Council of the U.K. Social Fund.

{}

\begin{appendix}

%\begin{figure*}
%\begin{centre}
%\begin{tabular}[b]{c}
%    \epsfig{file=../plots/Herschel250mic_CSO.eps, width=17cm,angle=0}\\
%\end{tabular}
%\caption{
%Colorscale and grey contours: field of J041757 observed with Herschel at 250 micron. Contours are $0.07$, $0.1$, to $0.32$ in steps of 0.02. Blue contours are the CSO 350 micron emission observed by our group. Note the striking agreement between all the CSO features and the Herschel features. Note the stream joining the source NE with B213, where J041757 is.
%}   
%\label{fcsocompact}
%\end{centre}
%\end{figure*}

%\begin{figure}
%\begin{centre}
%\begin{tabular}[b]{ccc}
%    \epsfig{file=../plots/Herschel350mic_CSO_6cm.eps, width=6cm,angle=0}\\
%    \epsfig{file=../plots/Herschel500mic_CSO_6cm.eps, width=6cm,angle=0}\\
%\end{tabular}
%\caption{
%Colorscale and grey contours: field of J041757 observed with Herschel at 250 micron. Contours are $0.07$, $0.1$, to $0.32$ in steps of 0.02. Blue contours are the CSO 350 micron emission observed by our group. Note the striking agreement between all the CSO features and the Herschel features. Note the stream joining the source NE with B213, where J041757 is.
%Bottom-right: Colorscale and grey contours: field of J041757 observed with Herschel at 250 micron. Contours are $0.07$, $0.1$, to $0.32$ in steps of 0.02. Blue contours are the CSO 350 micron emission observed by our group. Red contours: 13CO(1--0) integrated from 6.5 to 7.5 km/s (where the N2H+ is detected).
%}   
%\label{fj041757herschel}
%\end{centre}
%\end{figure}

\section{Photometry for J042118 and J041757-B}

In this Appendix we give the photometry for J042118 and J041757-B, from visible wavelengths up to the upper limit set at 1.2~mm with MAMBO-II. To compile the photometry we used the Virtual Observatory Spectral Energy Distribution Analyser (VOSA, Bayo \et\ 2008). 
See Hewett \et\ (2006) and Lucas \et\ (2008) for a review of the United Kingdom Infrared Digital Sky Survey.
For the case of J041757-B, we measured again the flux at 24~\mum\ to take into account extended emission (neglected in our previous measurement given in Barrado \et\ 2009).

\begin{table}
\caption{Photometry for J042118}
\begin{center}
{\small
\begin{tabular}{cccl}
\noalign{\smallskip}
\hline\noalign{\smallskip}
$\lambda$
&$S_\nu$
&$\sigma$
\\
($\mu$m)
&(mJy)
&(mJy)
&Instrument
\\
\noalign{\smallskip}
\hline\noalign{\smallskip}
0.89 &$<0.0142$ 	&-		&SDSS-z\\ 
1.03 &$<0.0130$  	&-		&UKIDSS-Y\\ 
1.25 &0.0165 		&0.0003 	&UKIDSS-J\\ 
2.20	&0.0453  		&0.0009 	&UKIDSS-K\\
3.4	&0.196		&0.010	&WISE\\
3.6	&0.249  		&0.004	&Spitzer/IRAC\\
4.5	&0.439  		&0.006	&Spitzer/IRAC\\
4.6	&0.435  		&0.023	&WISE\\
5.8	&0.674  		&0.010	&Spitzer/IRAC\\
8.0	&1.093  		&0.016	&Spitzer/IRAC\\
12	&1.13		&0.33	&WISE\\
24	&2.73  		&0.07	&Spitzer/MIPS\\
70	&$<33.6$  &$<4\sigma$	&Spitzer/MIPS\\
350   &46     		&9         	&CSO\\
1200	&$<4.0$ 		&$<4\sigma$	&IRAM\,30\,m\\
\hline
\end{tabular}
%\begin{list}{}{}
%\item[$^\mathrm{a}$] Positions corresponding to the intensity peak.
%\end{list}
}
\end{center}
\label{tsedj042118}
\end{table}

\begin{table}
\caption{Photometry for J041757-B}
\begin{center}
{\small
\begin{tabular}{cccl}
\noalign{\smallskip}
\hline\noalign{\smallskip}
$\lambda$
&$S_\nu$
&$\sigma$
\\
($\mu$m)
&(mJy)
&(mJy)
&Instrument
\\
\noalign{\smallskip}
\hline\noalign{\smallskip}
0.75 &0.0034 		&0.0001	&CFHT-i\\ 
0.90 &0.0077 		&0.0002	&CFHT-z\\ 
1.03 &$<0.0130$  	&-		&UKIDSS-Y\\ 
1.25 &0.0360 		&0.0006 	&CAHA-J\\ 
1.65 &0.0680 		&0.0012 	&CAHA-H\\ 
2.17	&0.127 		&0.002 	&CAHA-K\\
3.6	&0.295  		&0.008	&Spitzer/IRAC\\
4.5	&0.441  		&0.009	&Spitzer/IRAC\\
5.8	&0.491  		&0.010	&Spitzer/IRAC\\
8.0	&0.736  		&0.011	&Spitzer/IRAC\\
12	&1.79		&0.15	&WISE\\
22	&6.5			&1.0		&WISE\\
24	&7.5  		&1.0		&Spitzer/MIPS\\
70	&$<33.6$  &$<4\sigma$	&Spitzer/MIPS\\
350   &165     		&9         	&CSO\\
870	&$<80$		&$<4\sigma$	&APEX\\
1200	&$<2.9$ 		&$<4\sigma$	&IRAM\,30\,m\\
36000&0.19		&0.11	&VLA\\
60000&0.20		&0.05	&VLA\\
\hline
\end{tabular}
%\begin{list}{}{}
%\item[$^\mathrm{a}$] Positions corresponding to the intensity peak.
%\end{list}
}
\end{center}
\label{tsedj041757b}
\end{table}

%\subsection{Comparison of J041757-B properties to low-mass YSOs: a scaled-down version?}

%\begin{figure*}
%\begin{centre}
%\begin{tabular}[b]{cc}
% \epsfig{file=../plots/Shirley_07_Fig2a_J041757.eps,  width=7cm, angle=0} &
%    \epsfig{file=../plots/Shirley_07_Fig2d_J041757.eps, width=7cm,angle=0}\\
% \epsfig{file=../plots/Bontemps_96_Fig1_J041757.eps, width=7cm, angle=0}&
% \epsfig{file=../plots/Bontemps_96_Fig5_J041757.eps,  width=7cm, angle=0} \\
%\end{tabular}
%\caption{Correlations for the cm, outflow, Menv and Lbol for low-mass YSOs (Bontemps et al. 1996, Shirley et al. 2007), with the corresponding values for J041757 overplot on them, showing that the properties of J041757 are consistent with a scaled-down version of low-mass star formation, if J041757 is in Taurus.
%}
%\label{fscaledown}
%\end{centre}
%\end{figure*}

%Interesting papers from the literature:
%590]  arXiv:1112.3653 [pdf, ps, other]
%Spitzer IRAC identification of Herschel-ATLAS SPIRE sources
%Sam Kim, Julie L. Wardlow, Asantha Cooray, submitted to ApJ

\end{appendix}

\end{document}